\pdfoutput=1
\documentclass[lettersize,journal]{IEEEtran}
\usepackage{amsmath,amsfonts}
\usepackage{algorithmic}
\usepackage{algorithm}
\usepackage{array}
\usepackage[caption=false,font=normalsize,labelfont=sf,textfont=sf]{subfig}
\usepackage{textcomp}
\usepackage{stfloats}
\usepackage{url}
\usepackage{verbatim}
\usepackage{graphicx}
\usepackage{cite}
\usepackage{bm}
\usepackage{mathtools}
\usepackage{amssymb}
\usepackage{amsthm}

\newtheorem{lemma}{Lemma}

\usepackage{threeparttable}
\usepackage{booktabs}  
\usepackage{makecell}   
\usepackage{multirow}  
\usepackage{enumerate}
\usepackage{makecell} 
\usepackage{caption}
\usepackage{stfloats}
\usepackage{subeqnarray}

\usepackage{cases}

\begin{document}

\title{Flexible Rate-Splitting for Joint Unicast and Multi-Group Multicast Transmission in RIS-Assisted mmWave Networks}
\author{Hui Chen\textsuperscript{1}, Hongcheng Zhuang\textsuperscript{1,2}, Ahmed Badawy\textsuperscript{3}, Yanqun Tang\textsuperscript{1}\\
	{\normalsize \textsuperscript{1}School of Electronics and Communication Engineering,} 
	{\normalsize Sun Yat-sen University, Shenzhen 518107, China}\\
	{\normalsize \textsuperscript{2}Department of Broadband Communication, Peng Cheng Laboratory, Shenzhen 518055, China}\\
	{\normalsize \textsuperscript{3}Department of Computer Science and Engineering, Qatar University, Doha, Qatar}\\
	{\normalsize chenh525@mail2.sysu.edu.cn, zhuanghch@mail.sysu.edu.cn, badawy@qu.edu.qa, tangyq8@mail.sysu.edu.cn}
	\thanks{This work has been submitted to the IEEE for possible publication.  Copyright may be transferred without notice, after which this version may no longer be accessible.}
}
\maketitle

\begin{abstract}
Joint unicast and multi-group multicast transmission with RIS and RSMA is a promising enabler for 6G services. However, existing RSMA schemes for such scenarios split only unicast messages while leaving multicast messages intact, limiting the degree of freedom of interference management. To this end, we propose a joint rate splitting framework that splits both unicast and multicast information and two RSMA schemes. The common-common fusion (CCF-RSMA) scheme encodes the unicast common part into the global multicast common stream, while the private-common fusion (PCF-RSMA) scheme merges it with the group-specific multicast private part. For each scheme, we formulate energy efficiency (EE) maximization problems under both perfect and imperfect channel state information, and jointly optimize active beamforming, RIS phase shifts and rate allocation parameters. Simulation results demonstrate that the proposed schemes significantly outperform the comparative schemes in terms of EE, thereby proving the effectiveness of the proposed framework. Moreover, CCF-RSMA is more favorable in scenarios with larger groups and higher unicast QoS demands, whereas PCF-RSMA is better suited for scenarios with smaller groups and higher multicast QoS.
\end{abstract}


\section{Introduction}
Emerging applications in fifth-generation (5G) and beyond wireless networks necessitate simultaneous unicast and multi-group multicast transmissions \cite{intro1}. For instance, immersive augmented/virtual reality services can be generalized to multi-group multicast scenarios, where different virtual scenes are multicast to distinct user groups while personalized interactions are delivered via unicast. Such joint unicast–multicast transmission imposes stringent data rate requirements. To address this, the millimeter wave (mmWave) band is considered a key enabler due to its abundant available bandwidth\cite{intro2}. However, mmWave channels suffer from severe path loss and poor penetration, limiting their coverage and reliability. Reconfigurable Intelligent Surfaces (RIS) offer a promising way to enhance mmWave coverage and system performance by intelligently manipulating the wireless propagation environment\cite{intro3,intro4}. Thus, RIS-aided mmWave communication emerges as an efficient solution to support joint unicast-multicast transmission.
 
However, the joint unicast–multicast transmission introduces an extremely intricate multi-user interference structure, comprising interference among unicast users, among multicast groups, and between unicast and multicast transmissions. Moreover, multicast rates are limited by the weakest user within the group, together constraining the achievable rate improvement. Consequently, a multiple access scheme with built-in interference management is essential. Conventional space division multiple access (SDMA) requires sufficient spatial degrees of freedom and near-orthogonal user channels, limiting ihuibuhts effectiveness in overloaded or strong-interference scenarios\cite{intro5}. And power-domain layered division multiplexing non-orthogonal multiple access (LDM-NOMA) mandates decoding the multicast signal first at high power, restricting the flexibility of power allocation\cite{intro6}. In comparison, rate splitting multiple access (RSMA) offers a more flexible framework\cite{intro7}. Specifically, the transmitter adopts rate splitting (RS) to divide information into common and private parts, while the receiver uses SIC to decode part of the interference and treats the remainder as noise. This unlocks degrees of freedom in power allocation and interference management.

Recent studies have explored the potential of RSMA in mixed service scenarios. For unicast transmission with a single multicast group, the authors in \cite{intro8} propose 1-RSMA and generalized-RSMA schemes that split only unicast messages to trade off complexity and performance. The same 1-RSMA approach is adopted in \cite{intro9} for low earth orbit satellite networks, where the multicast message is merged into the common stream while unicast messages are split into common and private parts. For scenarios involving multiple multicast groups, \cite{intro10} investigates a STAR-RIS-assisted user grouping scheme based on joint services and channel conditions, where each group is assigned a specific number of subcarriers and 1-RSMA is applied within each group. The work in \cite{intro11} further extends RS to a multibeam satellite system that jointly delivers broadcast, multicast, and unicast services. In this scheme, each unicast message is split into three parts, which are respectively combined with the broadcast message, the multicast message, or transmitted as a private stream.

In summary, existing RSMA schemes focus on only splitting and optimizing unicast messages in mixed service scenarios. In scenarios with unicast and a single multicast group, treating the complete multicast information as a common stream is natural\cite{intro8,intro9}. However, this convention has been uncritically extended to multi-group multicast\cite{intro10,intro11}, unnecessarily restricting the full flexibility of RSMA to unicast messages. Critically, removing this unexamined convention unlocks a new design dimension: splitting  both unicast and multicast messages, which enables more comprehensive interference coordination across both traffic types. To explore this new design space, we propose a fully cooperative RSMA framework that simultaneously splits both unicast and multicast messages into common and private parts. With multicast splitting enabled, the design space of RSMA is no longer predetermined. This raises a previously unexplored architectural question: how should the common and private parts derived from unicast and multicast be structured? To answer this, the framework is implemented via two schemes. The first, termed common-common fusion RSMA (CCF-RSMA), encodes all common parts into a single super-common stream. The second, termed private-common fusion RSMA (PCF-RSMA), forms a per-group common stream by jointly encoding the private part of the multicast message and the common parts of the unicast messages intended for users within the same multicast group. Within this framework, we jointly optimize the messages split ratios, beamforming vectors, and RIS phase shifts to achieve flexible interference management.

However, applying these schemes to practical RIS-assisted mmWave transmission faces the critical challenge of acquiring channel state information (CSI). This difficulty stems primarily from the passive nature of RIS, the sparsity of mmWave channels and the resulting hardship in estimating the high-dimensional cascaded channels, compounded by the limited pilot overhead in practical systems \cite{intro12, intro13,intro13-1}. Consequently, evaluating the performance benefits under imperfect CSI is vital. While RSMA exhibits robustness to CSI imperfections in multi-user interference management \cite{intro14,intro14-1,intro14-2}, in the studied RIS-aided joint unicast-multicast transmission, the robustness issue is more complex due to the performance of multicast groups bounded by the worst user within each group. 

To assess the overall system efficiency of the proposed schemes, we adopt energy efficiency (EE) as the primary performance metric, evaluating it under both ideal and practical CSI conditions. This importance stems not only from the general demand for green communications\cite{intro15,intro16}, but from the schemes' design, i.e., the message splitting ratios for the two types of services serve as novel optimization variables. These ratios, together with beamforming vectors and RIS phase shifts, jointly determine the trade-off between achievable rate and power consumption. Accordingly, we formulate an EE maximization problem to jointly optimize all these parameters, which enables a systematic evaluation of the proposed schemes' performance. The main contributions are summarized as follows.
\begin{itemize}
	\item  We propose a general RSMA framework that, for the first time, extends message splitting to both unicast and multicast for joint unicast and multi-group multicast scenarios. This framework is realized through two novel schemes, CCF-RSMA and PCF-RSMA. Each scheme is based on a distinct philosophy for fusing common and private parts, thereby enabling flexible and coordinated cross-service interference management.

    \item We formulate dedicated EE maximization problems for both CCF-RSMA and PCF-RSMA, each jointly optimizing message split ratios, active beamforming and RIS phase shifts under both perfect and imperfect CSI. To the best of our knowledge, this is the first work to jointly address 1) the hierarchical interference from joint unicast-multicast message splitting, 2) the multicast group rate bottleneck, and 3) RIS-assisted worst-case CSI uncertainties within a unified optimization paradigm.
    
    \item To demonstrate the practical tractability of the proposed EE maximization problems, we develop a customized algorithmic solutions. For the perfect CSI case, the main challenge lies in the non-convex coupling between beamforming and RIS phase shifts. We tackle this via a block coordinate ascent ((BCA) framework, where each subproblem is handled by successive convex approximation (SCA). Specifically, for the RIS phase-shift subproblem, we design a penalty-based SCA method that iteratively relaxes the unit-modulus constraints, ensuring monotonic convergence to a stationary point.
    
    \item We further extend the algorithmic framework to imperfect CSI with bounded uncertainties. This introduces additional challenges in modeling worst-case robustness. We employ the S-procedure to convert uncertainty constraints into linear matrix inequalities (LMIs), and solve the resulting robust optimization via a semidefinite relaxation (SDR) approach enhanced by a sequential rank-one constrained relaxation (SROCR) technique. This enables robust EE optimization in RIS-aided RSMA systems with joint unicast and multi-group multicast transmissions under imperfect CSI.
    
   \item Simulations validate the effectiveness of the proposed framework and reveal that the optimal fusion scheme depends on system parameters. Specifically, CCF-RSMA, which prioritizes global interference coordination, exceeds with larger groups and low multicast QoS requirements. In contrast, PCF-RSMA, designed for flexible intra-group rate adaptation, is preferable when groups are smaller with high multicast QoS demands.
\end{itemize}

The rest of this paper is organized as follows. The system model, channel model, and  two information splitting schemes are introduced in Section II. Section III investigates the EE maximization design for active and passive beamforming under the perfect CSI assumption. Section IV extends the design to a bounded CSI error model for robust beamforming. Section V describes the complexity of the proposed algorithms. Section VI presents simulation results to evaluate the effectiveness of the algorithm. Finally, the paper concludes in Section VII. Moreover, the main symbols we adopt are listed in Table \ref{table1}.

\paragraph*{Notations:}
Boldface letters denote matrices/vectors; $\mathbf{I}_M$ is the $
M\times M$ identity matrix; $(\cdot)^H$, Tr \{$\cdot$\}, and diag $\{\cdot\}$  represent the Hermitian 
transpose, trace, and diagonalization; $\boldsymbol{A} \succeq \boldsymbol{0}$ indicates positive semidefinite; 
$\|\cdot\|$ is the Euclidean norm; $\Re\{\cdot\}$ and $\mathcal{O}(\cdot)$ denote real part and computational complexity. Table \ref{table1} summarizes the key symbols and functions used throughout the paper.

\begin{table}[h]
	\caption{List of Symbols and Functions}
	\centering
	\begin{tabular}{|l|l|}
		\hline Notation & Definition \\
		\hline $\boldsymbol{d}_k$ & Direct channel of BS-user $k$  \\
		\hline $\boldsymbol{h}_k$ & Cascaded channel of RIS-user $k$ \\
		\hline $\boldsymbol{H}$ & Cascaded channel of BS-RIS \\
		\hline $\boldsymbol{\Phi}(\boldsymbol{\phi})$ & Matrix (vector) of RIS phase shifts \\
		\hline $W_{p,k}^u$ & Private part of the unicast message for user $k$ \\
		\hline $W_{c,k}^u$ & Common part of the unicast message for user $k$ \\
	    \hline $W_{p,g}^m$ & Private part of the multicast message for group $g$ \\ 
	    \hline $W_{c,g}^m$ & Common part of the multicast message for group $g$\\ 
	    \hline $\boldsymbol{w}_{c}$ & Beamforming vector of common stream\\
	    \hline $\boldsymbol{w}_{p,k}^u$ & Beamforming vector of the $k$-th private unicast stream \\
	    \hline $\boldsymbol{w}_{p,g}^m$ & \makecell{Beamforming vector of the $g$-th private multicast stream }\\
	    \hline $\acute{\boldsymbol{w}}_{c,g}^{gs}$ & \makecell{Beamforming vector of the $g$-th group-specific \\ common stream }\\
	    \hline $C_{c,k}^u,C_{c,g}^m$ & \makecell{Common sub-message allocation ratios }\\
	    \hline $\acute{C}_{c,k}^{gs,u},\acute{C}_{c,g}^{gs,m}$ &\makecell{Group-specific common sub-message allocation  ratios} \\
	    \hline $f(\cdot)$ &Convex approximation for coupled terms\\
	    \hline $\mathcal{L}_{lb}(\cdot)$ & LMI for lower bound (matrix arguments)\\
	    \hline $\mathcal{L}_{ub}(\cdot)$ &LMI for upper bound (matrix arguments)  \\
	    \hline $\mathcal{F}_{lb}(\cdot)$ &LMI for lower bound (vector arguments)  \\  
	    \hline $\mathcal{F}_{ub}(\cdot)$ &LMI for upper bound (vector arguments) \\
		\hline
	\end{tabular}
	\smallskip
	\vspace{2pt}
	{\footnotesize Note: All PCF-RSMA-specific symbols are marked with an acute accent $(\acute{})$.}
\label{table1}
\end{table}
\vspace{-15pt}
\section{SYSTEM MODEL}
\begin{figure}[htbp]
		\centering
		\includegraphics[width=0.4\textwidth]{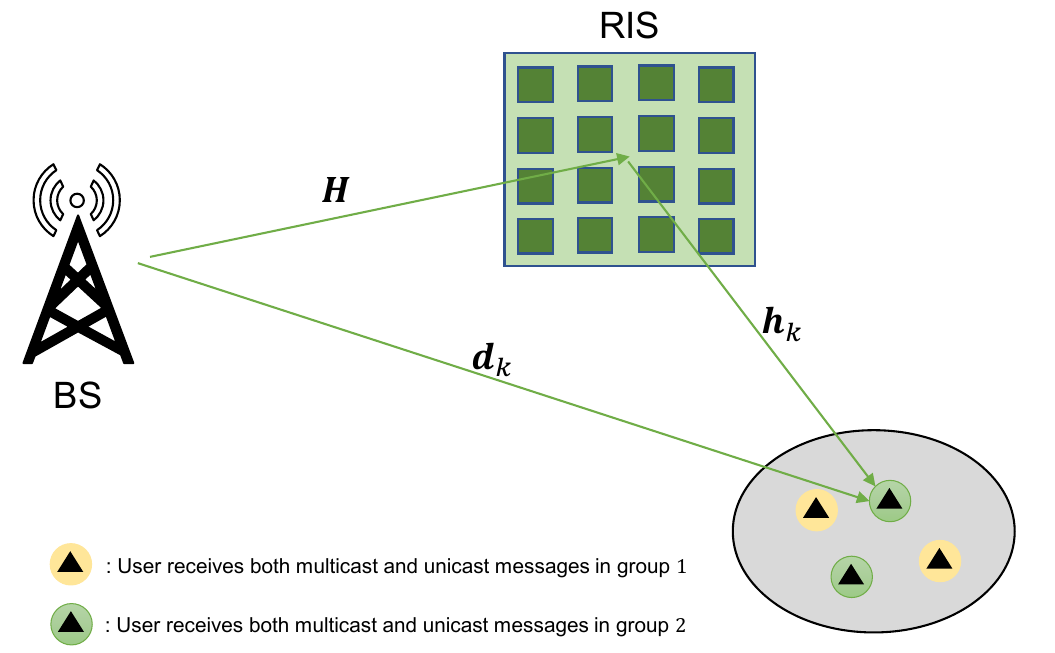}
	\caption{RIS-assisted mmWave systems for joint unicast and multigroup multicast transmission.}
	\label{fig1}
\end{figure}
We consider a RIS-assisted mmWave communication that supports joint unicast and multigroup multicast transmission, as illustrated in Fig. \ref{fig1}. Specifically, a base station (BS) with $M$ antennas
serves $K$ single-antenna users, denoted by $\mathcal{K} = \{1, \cdots, K\}$, with the assistance of an $N$-element RIS. Each user is served with both a dedicated unicast message and a common multicast message corresponding to its group membership. Users sharing the same multicast message are grouped together, forming $G$ multicast groups indexed by $\mathcal{G} = \{1, \cdots, G\}$, where $\mathcal{K}_g$ denotes the set of users in the $g$-th group. These group sets constitute a partition of the total user set $\mathcal{K}$, i.e., $\mathcal{K}=\cup_{g=1}^G\mathcal{K}_g$ and $\mathcal{K}_i\cap\mathcal{K}_j=\emptyset$ for all $i\neq j$. 
\vspace{-10pt}
\subsection{Channel model}
We adopt the Saleh-Valenzuela channel model to characterize mmWave channels due to its sparse nature \cite{intro27,intro28}. Specifically, the direct link $\boldsymbol{d}_k\in\mathbb{C}^{M\times 1}$ between the BS and user $k$ is modeled as
\begin{equation}\label{eq1}
	\boldsymbol{d}_k= \lambda_B\lambda_U\sqrt{M/L_{k,BU}}\sum\nolimits_{l=1}^{L_{k,BU}}\alpha_{l,k} \boldsymbol{a}_{B}\left(\phi_{k,l}^{\text{AoD}}, \theta_{k,l}^{\text{AoD}}\right),
\end{equation}
where $\lambda_B$ and $\lambda_U$ are the transmit and receive gain, respectively. $\alpha_{l,k}$ and $L_{k,BU}$ denote the complex gain and the number of mmWave propagation paths, respectively. $\phi_{k,l}^{\text{AoD}}$ and $\theta_{k,l}^{\text{AoD}}$ represent the azimuth and elevation angles of departure (AoD) of the path, respectively. 

The BS-RIS channel $\boldsymbol{H}\in\mathbb{C}^{N\times M}$ and the RIS-user $k$ channel $\boldsymbol{h}_k\in\mathbb{C}^{N\times 1}$ are given as follows, respectively
\begin{equation}\label{eq2}
	\boldsymbol{H}=\lambda_B\lambda_R\sqrt{\frac{MN}{L_{BR}}}\sum_{l=0}^{L_{BR}-1} \beta_{l} \boldsymbol{a}_{R}\left(\phi_{l}^{\text{AoA}}, \theta_{l}^{\text{AoA}}\right) \boldsymbol{a}_{B}^{H}\left(\phi_{l}^{\text{AoD}}, \theta_{l}^{\text{AoD}}\right),
\end{equation}
\begin{equation}\label{eq3}
	\boldsymbol{h}_k=\lambda_R\sqrt{N/L_{k,RU}}\sum\nolimits_{l=0}^{L_{k,RU}-1} \beta_{l,k} \boldsymbol{a}_{B}\left(\phi_{k,l}^{\text{AoD}}, \theta_{k,l}^{\text{AoD}}\right),
\end{equation}
where $\lambda_R$ is the receive gain and $\beta_{l} (\beta_{l,k})$ are the complex gains. $L_{BR}$ and $L_{k,RU}$ are the number of mmWave channel paths for cascaded link BS-RIS and RIS-user $k$, respectively. $\phi_{l}^{\text{AoA}}$ and $\theta_{l}^{\text{AoA}}$ stand for the azimuth and elevation angles of arrival (AoA) of the path $l$, respectively. The index $l=0$ in \eqref{eq1}-\eqref{eq3} denotes the line-of-sight (LoS) path, while $l>0$ corresponds to non-line-of-sight (NLoS) paths.

In the above, $\boldsymbol{a}_{B}(\cdot)$ and $\boldsymbol{a}_{R}(\cdot)$ denote the array response vectors of the BS and the RIS, respectively. Both the BS and RIS are equipped with uniform planar arrays (UPA). The expression of the UPA response vector is given by:
\begin{equation}
	\begin{aligned}
		\boldsymbol{a}_{\text{UPA}}  \left(\phi, \theta\right) 
		= & \frac{1}{\sqrt{N_hN_v}}\big[1, \ldots, e^{j \frac{2 \pi d}{\lambda}\left(i_h \sin \phi \sin\theta+i_v \cos\theta\right)} \\
		& \ldots, e^{j \frac{2 \pi d}{\lambda}\left(\left(N_h-1\right) \sin\phi \sin \theta+\left(N_v-1\right) \cos\theta\right)}\big]^{\mathrm{T}},
	\end{aligned}
\end{equation}
where $d$ and $\lambda$ denote the inter-antenna spacing and the signal wavelength, respectively. $N_h$ and $N_v$ are the numbers of horizontal and vertical antennas, with antenna indices $i_h \in [0, N_h-1]$ and $i_v \in [0, N_v-1]$, respectively.
\vspace{-10pt}
\subsection{Transmission Schemes}
We consider a 2-layer RSMA framework for joint unicast and multi-group multicast transmissions. Following the SIC principle that information shared by more users should be decoded first\cite{intro8}, each user sequentially decodes: 1) a global common stream, 2) a group-level common stream, and 3) its own unicast private stream. With this decoding order fixed, the only remaining design freedom is where to place the common part of each unicast message. Given the two-layer structure, the common part can only be allocated to one of two possible layers: the global layer shared by all users or the group layer shared only within its multicast group. This leads to two distinct architectures:
\begin{enumerate}[1)]
	\item CCF-RSMA: placing the unicast common part in the global layer, merging it with the common parts of all multicast groups;  
	\item PCF-RSMA: placing it in the group layer, merging it with the private part of its own multicast group.
\end{enumerate} 
Next, we elaborate on the two schemes in detail under the perfect CSI availability.

\subsubsection{CCF-RSMA}
 \begin{figure}[htbp]
	\centering
	\includegraphics[width=0.4\textwidth]{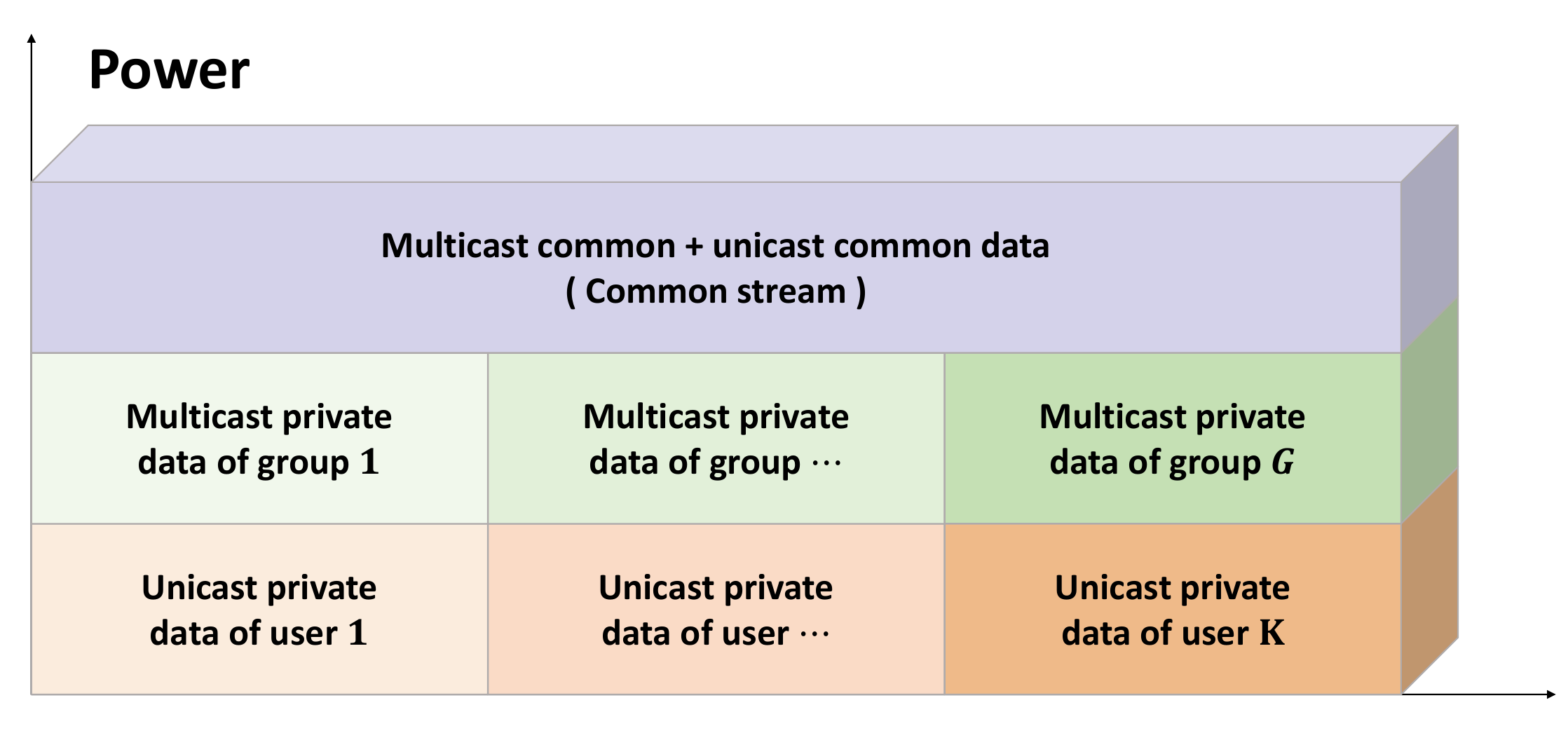}
	\caption{Data types of 2-layer CCF-RSMA.}
	\label{fig2}
\end{figure}
In the proposed 2-layer CCF-RSMA scheme, we partition each unicast message $W_k^u$ into a private part $W_{p,k}^u$ and a common part $W_{c,k}^u$. Similarly, the multicast message $W_g^m$ for group $g$ is also split into a common part $W_{c,g}^m$ and a private part $W_{p,g}^m$. All common parts $\{W_{c,1}^u,\cdots,W_{c,K}^u\}$ and $\{W_{c,1}^m,\cdots,W_{c,G}^m\}$ are jointly encoded into a common stream $s_c$. The private parts $W_{p,k}^u$ and $W_{p,g}^m$ are independently encoded into their respective private streams $s_{p,k}^u$ and $s_{p,g}^m$, as shown in Fig. \ref{fig2}. The resulting stream vector is composed of all streams, denoted as $\boldsymbol{s}=[s_c,s_{p,1}^m,\cdots,s_{p,G}^m,s_{p,1}^u,\cdots,s_{p,K}^u]$, and we assume $\mathbb{E}\{\boldsymbol{s}\boldsymbol{s}^H\}=\boldsymbol{I}$. Denote the forming matrix $\boldsymbol{W}=[\boldsymbol{w}_c,\boldsymbol{w}_{p,1}^m,\cdots,\boldsymbol{w}_{p,G}^m,\boldsymbol{w}_{p,1}^u,\cdots,\boldsymbol{w}_{p,K}^u]$, where $\boldsymbol{w}_c\in\mathbb{C}^{M\times1}$, $\boldsymbol{w}_{p,g}^m\in\mathbb{C}^{M\times1}$ and $\boldsymbol{w}_{p,k}^u\in\mathbb{C}^{M\times1}$ are the beamforming vectors for the symbol $s_c$, $s_{p,g}^m$ and $s_{p,k}^u$, respectively. The transmitted signal vector from the BS is thus given by
 \begin{equation}
 	\begin{aligned}
 		\boldsymbol{x}=&\boldsymbol{w}_{c} s_{c}+\sum\nolimits_{g\in\mathcal{G}} \boldsymbol{w}_{p,g}^m s_{p,g}^m+\sum\nolimits_{k\in\mathcal{K}}\boldsymbol{w}_{p,k}^u s_{p,k}^u.
 	\end{aligned}
 \end{equation}
 Then, the received signal at the user $k$ can be presented as
 \begin{equation}
 	\begin{aligned}
 		y_k=&\underbrace{(\boldsymbol{h}_k^H\boldsymbol{\Phi}\boldsymbol{H}+\boldsymbol{d}_k^H)\boldsymbol{w}_{p,g_k}^m s_{p,g_k}^m}_{\text {desired multicast private signal}}+\underbrace{(\boldsymbol{h}_k^H\boldsymbol{\Phi}\boldsymbol{H}+\boldsymbol{d}_k^H)\boldsymbol{w}_{p,k}^u s_{p,k}^u}_{\text {desired unicast private signal}}\\
 		&+\underbrace{(\boldsymbol{h}_k^H\boldsymbol{\Phi}\boldsymbol{H}+\boldsymbol{d}_k^H)\boldsymbol{w}_{c} s_{c}}_{\text {desired common signal}}+\underbrace{\sum_{g\neq g_k}(\boldsymbol{h}_k^H\boldsymbol{\Phi}\boldsymbol{H}+\boldsymbol{d}_k^H)\boldsymbol{w}_{p,g}^m s_{p,g}^m}_{\text {inter-group interference}}\\
 		&+\underbrace{\sum\nolimits_{j\neq k}(\boldsymbol{h}_k^H\boldsymbol{\Phi}\boldsymbol{H}+\boldsymbol{d}_k^H)\boldsymbol{w}_{p,j}^u s_{p,j}^u}_{\text {inter-user interference}}+n_k,k\in\mathcal{K},
 	\end{aligned}
 \end{equation}
 where $g_k\in\mathcal{G}$ denotes the index of the multicast group to which user $k$ is assigned, and $n_k\sim (0,\sigma^2)$ represents the additive
 white Gaussian noise (AWGN) for user $k$. The matrix $\boldsymbol{\Phi}=\text{diag}(\boldsymbol{\phi})=\text{diag}\left(\left[e^{j\theta_1}, e^{j\theta_2}, \cdots, e^{j\theta_{N}}\right]\right) \in \mathbb{C}^{N \times N}$ denotes reflection coefficient of the RIS, where $\theta_n\in[0,2\pi]$. For brevity, we denote by $\boldsymbol{F}_k = \text{diag}(\boldsymbol{h}_k^H) \boldsymbol{H}$ the cascaded channel from the BS to user $k$ via the RIS.
 
The SIC decoding order follows the rule that data streams are decoded in descending order of the size of their intended user sets\cite{intro8}. Therefore, each user first decodes the common stream $s_c$ by treating all the private streams as noise. The achievable rate for user $k$ to decode $s_c$ is given by
\begin{equation}
 	R_{c,k}=\log_2\bigg(1+\frac{\left|(\boldsymbol{\phi}^H\boldsymbol{F}_k+\boldsymbol{d}_k^H)\boldsymbol{w}_c\right|^2}{I_{c,k}+\sigma_k^2}\bigg),
 \end{equation}
 where $I_{c,k}=\sum\nolimits_{g\in\mathcal{G}}\left|(\boldsymbol{\phi}^H\boldsymbol{F}_k+\boldsymbol{d}_k^H) \boldsymbol{w}_{p,g}^m\right|^2+\sum\nolimits_{j\in\mathcal{K}}\left|(\boldsymbol{\phi}^H\boldsymbol{F}_k+\boldsymbol{d}_k^H) \boldsymbol{w}_{p,j}^u\right|^2$ aggregates the interference from all multicast and unicast private streams, which are treated as noise during the decoding of the common stream $s_c$. To ensure that all users can successfully decode $s_c$, the actual transmission rate $R_c$ of the common stream must not exceed the achievable rate $R_{c,k}$, i.e., $R_{c}=\min_{\{k\in\mathcal{K}\}}R_{c,k}$. In CCF-RSMA, the common stream $s_c$ carries the common parts of all unicast messages as well as the multicast messages. Therefore, the total common rate $R_c$ is the sum of the respective rates allocated to these components, i.e.,$R_c=\sum_{k=1}^{K}C_{c,k}^u+\sum_{g=1}^{G}C_{c,g}^m$, where $C_{c,k}^u$ is the portion allocated to the common sub-message of the $k$-th unicast user, and $C_{c,g}^m$ is the rate allocated to the common sub-message of the $g$-th multicast group. These values are determined by solving the optimization problem in the next section.
 
After successfully decoding and removing the common stream $s_c$, user $k$ proceeds to decode its designated multicast private stream from group $g_k$. The achievable rate for user $k$ to decode multicast private stream can be expressed as
 \begin{equation}
 	R_{p,g}^m=\min_{k\in\mathcal{K}_{g}}\log_2\bigg(1+\frac{\left|(\boldsymbol{\phi}^H\boldsymbol{F}_k+\boldsymbol{d}_k^H) \boldsymbol{w}_{p,g}^m\right|^2}{I_{p,k}^m+\sigma_k^2}\bigg),
 \end{equation}
 where $I_{p,k}^m=\sum_{g'\neq g}\left|(\boldsymbol{\phi}^H\boldsymbol{F}_k+\boldsymbol{d}_k^H)\boldsymbol{w}_{p, g'}^m\right|^2+\sum_{j=1}^K\left|(\boldsymbol{\phi}^H\boldsymbol{F}_k+\boldsymbol{d}_k^H)\boldsymbol{w}_{p,j}^u\right|^2$ captures the interference from the private streams of all other multicast groups and all unicast private streams.
 
Following the successful cancellation of $s_c$ and $s_{p,g_k}^m$, user $k$ decodes its dedicated unicast private stream with the following achievable rate
 \begin{equation}
 	R_{p,k}^u=\log_2\bigg(1+\frac{\left|(\boldsymbol{\phi}^H\boldsymbol{F}_k+\boldsymbol{d}_k^H)\boldsymbol{w}_{p,k}^u\right|^2}{I_{p,k}^u+\sigma_k^2}\bigg),
 \end{equation}
 where $I_{p,k}^u=\sum\nolimits_{g\neq g_k}\left|(\boldsymbol{\phi}^H\boldsymbol{F}_k+\boldsymbol{d}_k^H)\boldsymbol{w}_{p, g}^m\right|^2+\sum\nolimits_{j\neq k}\left|(\boldsymbol{\phi}^H\boldsymbol{F}_k+\boldsymbol{d}_k^H)\boldsymbol{w}_{p,j}^u\right|^2$ denotes the residual interference power, comprising the private streams of all other multicast groups and all other unicast users.
 
 Therefore, the overall achievable rate for unicast user $k$ and multicast group $g$ are respectively given by
 \begin{equation}
 	R_k^{u}=C_{c,k}^u+R_{p,k}^u,
 \end{equation}
 \begin{equation}
 	R_g^m=C_{c,g}^m+R_{p,g}^m.
 \end{equation}
 
The total system power consumption is represented as
 \begin{equation}
 	P_{total}=\frac{1}{\eta}(\Vert\boldsymbol{w}_c\Vert^2+\sum_{g=1}^G\Vert\boldsymbol{w}_{p,g}^m\Vert^2+\sum_{k=1}^K\Vert\boldsymbol{w}_{p,k}^u\Vert^2)+P_{cir},
 \end{equation}
 where $\eta$ is the power amplifier efficiency. The total circuit power is $P_{cir}=P_{BS}+P_{RIS}+KP_U$, with $P_{RIS}$ and $P_{U}$ denoting the static power of the RIS and per user. Since a fully digital beamforming architecture is not employed in this work, the number of radio frequency (RF) chains is set to one. Thus, the circuit power consumption at the mmWave BS can be decomposed as $P_{BS} = P_{BB} + P_{\text{RF}} + M(P_{PS} + P_{PA})$\cite{intro29}, where $P_{BB}$, $P_{RF}$, $P_{PS}$ and $P_{PA}$ are the power consumption of the baseband unit, RF chain, phase shifter and power amplifier, respectively.
 
\subsubsection{PCF-RSMA}
\begin{figure}[htbp]
	\centering
	\includegraphics[width=0.40\textwidth]{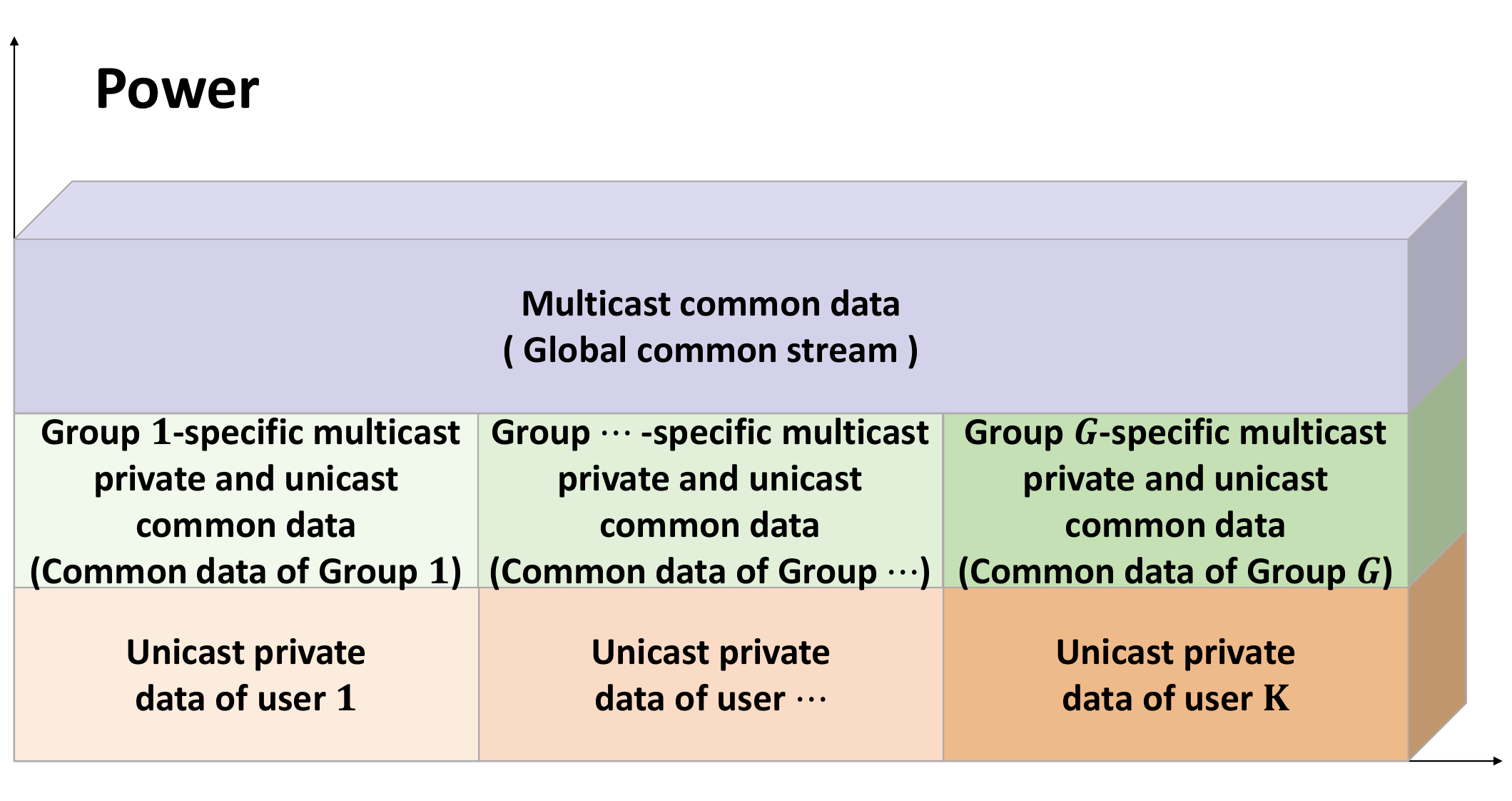}
	\caption{Data types of 2-layer PCF-RSMA.}
	\label{fig3}
\end{figure}
Within the 2-layer PCF-RSMA scheme, the global common stream $\acute{s}_c$ is a joint encoding of all common parts of multicast messages $\{\acute{W}_{c,1}^m,\cdots,\acute{W}_{c,G}^m\}$ to handle the inter-group interference. To manage intra-group interference localized within group $g$, the group-specific common stream $\acute{s}_{c,g}^{gs},g\in\mathcal{G}$ is formed by jointly encoding the private part of the group's multicast message and the common parts of the unicast messages for its members, i.e., $\acute{W}_{c,g}^{gs}=\{\acute{W}_{p,g}^m,\acute{W}_{c,1}^u,\cdots,\acute{W}_{c,\vert\mathcal{K}_g\vert}^u\}, g\in\mathcal{G}$. The private parts $\acute{W}_{p,k}^u$ are independently encoded into private streams $\acute{s}_{p,k}^u$. The complete stream vector at the BS is therefore $\acute{\boldsymbol{s}}=[\acute{s}_c,\acute{s}_{c,1}^{gs},\cdots,\acute{s}_{c,G}^{gs},\acute{s}_{p,1}^u,\cdots,\acute{s}_{p,K}^u]$, satisfying $\mathbb{E}\{\acute{\boldsymbol{s}}\acute{\boldsymbol{s}}^H\}=\boldsymbol{I}$. Furthermore, the stream vector $\acute{\boldsymbol{s}}^T$ is processed by the beamforming matrix $\acute{\boldsymbol{W}}=[\acute{\boldsymbol{w}}_c,\acute{\boldsymbol{w}}_{c,1}^{gs},\ldots,\acute{\boldsymbol{w}}_{c,G}^{gs}, \acute{\boldsymbol{w}}_{p,1}^u,\ldots,\acute{\boldsymbol{w}}_{p,K}^u]$. Here, $\acute{\boldsymbol{w}}_c\in\mathbb{C}^{M\times1}$, $\acute{\boldsymbol{w}}_{c,g}^{gs}\in\mathbb{C}^{M\times1}$ and $\acute{\boldsymbol{w}}_{p,k}^u\in\mathbb{C}^{M\times1}$ are the beamforming vectors corresponding to the streams $\acute{s}_c$, $\acute{\boldsymbol{s}}_{c,g}^{gs}$ and $\acute{s}_{p,k}^u$, respectively. Thus, the transmitted signal of the BS is
\begin{equation}
		\acute{\boldsymbol{x}}=\acute{\boldsymbol{w}}_{c} \acute{s}_{c}+\sum\nolimits_{g\in\mathcal{G}} \acute{\boldsymbol{w}}_{g}^{gs} \acute{s}_{g}^{gs}+\sum\nolimits_{k\in\mathcal{K}}\acute{\boldsymbol{w}}_{p,k}^u \acute{s}_{p,k}^u.
\end{equation}

The received singal at user $k$ is given by
\begin{equation}
	\begin{aligned}
		\acute{y}_k=&\underbrace{(\acute{\boldsymbol{\phi}}^H\boldsymbol{F}_k+\boldsymbol{d}_k^H)\acute{\boldsymbol{w}}_{g_k}^{gs} \acute{s}_{g_k}^{gs}}_{\text {desired group-specific signal}}+\underbrace{(\acute{\boldsymbol{\phi}}^H\boldsymbol{F}_k+\boldsymbol{d}_k^H)\acute{\boldsymbol{w}}_{p,k}^u \acute{s}_{p,k}^u}_{\text {desired unicast private signal}}\\
		&+\underbrace{(\acute{\boldsymbol{\phi}}^H\boldsymbol{F}_k+\boldsymbol{d}_k^H)\acute{\boldsymbol{w}}_{c} \acute{s}_{c}}_{\text {desired common signal}}+\underbrace{\sum\nolimits_{g\neq g_k}(\acute{\boldsymbol{\phi}}^H\boldsymbol{F}_k+\boldsymbol{d}_k^H)\acute{\boldsymbol{w}}_{g}^{gs} \acute{s}_{g}^{gs}}_{\text {inter-group interference}}\\
		&+\underbrace{\sum\nolimits_{j\neq k}(\acute{\boldsymbol{\phi}}^H\boldsymbol{F}_k+\boldsymbol{d}_k^H)\acute{\boldsymbol{w}}_{p,j}^u \acute{s}_{p,j}^u}_{\text {inter-user interference}}+\acute{n}_k.
	\end{aligned}
\end{equation}

User $k$ first decodes the global common stream from its received signal and then removes it via SIC. The achievable rate of decoding $\acute{s}_c$ at the user $k$ writes as
\begin{equation}
	\acute{R}_{c,k}=\log_2\bigg(1+\frac{\left|(\acute{\boldsymbol{\phi}}^H\boldsymbol{F}_k+\boldsymbol{d}_k^H)\acute{\boldsymbol{w}}_c\right|^2}{\acute{I}_{c,k}^c+\sigma_k^2}\bigg),\forall k \in\mathcal{K},
\end{equation}
where $\acute{I}_{c,k}^c = \sum_{g\in\mathcal{G}} |(\acute{\boldsymbol{\phi}}^H\boldsymbol{F}_k+\boldsymbol{d}_k^H)\acute{\boldsymbol{w}}_{c,g}^{gs}|^2 + \sum_{j\in\mathcal{K}} |(\acute{\boldsymbol{\phi}}^H\boldsymbol{F}_k+\boldsymbol{d}_k^H)\acute{\boldsymbol{w}}_{p,j}^u|^2$ represents the interference power. The corresponding common rate satisfies $\acute{R}_c=\min_{k\in\mathcal{K}} \acute{R}_{c,k} =\sum_{g\in\mathcal{G}}\acute{C}_{c,g}^m$, where $\acute{C}_{c,g}^m$ denoteS the portions allocated to the common part of the $g$-th group's message. After successfully decoding and removing $\acute{s}_c$, the achievable rate for group-specific common stream at user $k$ is expressed as
\begin{equation}
	\acute{R}_{c,k}^{gs}=\log_2\bigg(1+\frac{\left|(\acute{\boldsymbol{\phi}}^H\boldsymbol{F}_k+\boldsymbol{d}_k^H)\acute{\boldsymbol{w}}_{g}^{gs}\right|^2}{\acute{I}_{c,k}^{gs}+\sigma_k^2}\bigg),k\in \mathcal{K}_g,
\end{equation}
where $\acute{I}_{c,k}^{gs}=\sum_{g'\neq g}\left|(\acute{\boldsymbol{\phi}}^H\boldsymbol{F}_k+\boldsymbol{d}_k^H)\acute{\boldsymbol{w}}_{g'}^{gs}\right|^2+\sum_{k\in\mathcal{K}}\left|(\acute{\boldsymbol{\phi}}^H\boldsymbol{F}_k+\boldsymbol{d}_k^H)\acute{\boldsymbol{w}}_{p,k}^u\right|^2$ is the interference for user $k$ in group $g$. $\acute{s}_{g}^{gs}$ carries the private parts of the multicast message of group $g$ as well as the common parts of all unicast messages for users within this group. Therefore, the total group-specific common rate $\acute{R}_{c,g}^{gs}$ is the sum of the rates allocated to these components, i.e., $\acute{R}_{c,g}^{gs} = \min_{k\in\mathcal{K}_g} \acute{R}_{c,k}^{gs} = \sum_{k \in \mathcal{K}_g} \acute{C}_{c,k}^{gs,u} + \acute{C}_{p,g}^{gs,m}$, where $\acute{C}_{c,k}^{gs,u}$ and $\acute{C}_{p,g}^{gs,m}$ represent the portions allocated to the common part of user $k$'s unicast message and the private part of group $g$'s multicast message, respectively.

After successfully decoding and removing the target group-specific common stream, the achievable rate for unicast private stream at user $k$ is written as
\begin{equation}
	\acute{R}_{p,k}^u=\log_2\bigg(1+\frac{\left|(\acute{\boldsymbol{\phi}}^H\boldsymbol{F}_k+\boldsymbol{d}_k^H)\acute{\boldsymbol{w}}_{p,k}^u\right|^2}{\acute{I}_{p,k}^u+\sigma_k^2}\bigg),k\in \mathcal{K},
\end{equation}
where $\acute{I}_{p,k}^u=\sum_{g\neq g_k}\left|(\acute{\boldsymbol{\phi}}^H\boldsymbol{F}_k+\boldsymbol{d}_k^H)\acute{\boldsymbol{w}}_{g}^{gs}\right|^2+\sum_{j\neq k}\left|(\acute{\boldsymbol{\phi}}^H\boldsymbol{F}_k+\boldsymbol{d}_k^H)\acute{\boldsymbol{w}}_{p,j}^u\right|^2$.

The achievable rate for unicast user $k$, which combines its allocated common-message portions and private rate, is 
\begin{equation}
	\acute{R}_k^{u}=\acute{C}_{c,k}^{gs,u}+\acute{R}_{p,k}^u.
\end{equation}
The achievable rate for multicast group $g$, comprising parts from both the global common and group-specific common streams, is represented as
\begin{equation}
	\acute{R}_g^m=\acute{C}_{c,g}^m+\acute{C}_{p,g}^{gs,m},
\end{equation}
where the sum reflects that the multicast message is allocated across both the global common and the group-common layer.

Finally, the total system power consumption, accounting for both the transmit power and circuit power $\acute{P}_{cir}$, is denoted by
\begin{equation}
	\acute{P}_{total}=\frac{1}{\eta}(\Vert\acute{\boldsymbol{w}}_c\Vert^2+\sum_{g=1}^G\Vert\acute{\boldsymbol{w}}_{g}^{gs}\Vert^2+\sum_{k=1}^K\Vert\acute{\boldsymbol{w}}_{p,k}^u\Vert^2)+\acute{P}_{cir}.
\end{equation}

\section{PROPOSED JOINT BEAMFORMING DESIGN}
In this section, we aim to design joint active and passive beamforming scheme to maximize system EE under QoS and transmit power constraints for joint unicast and multi-group multicast transmissions. We consider both CCF-RSMA and PCF-RSMA schemes under the perfect CSI assumption.
\subsection{Joint Beamforming Design for CCF-RSMA}
For the CCF-RSMA scheme, the optimization problem to maximize EE can be formulated as
\begin{subequations}
	\begin{align}
		\mathcal{P}_1:&
		\max _{\{\boldsymbol{W},\boldsymbol{\phi},\boldsymbol{C}\}} \frac{\sum\nolimits_{k\in\mathcal{K}} R_{k}^u+\sum\nolimits_{g\in\mathcal{G}} R_g^m}{P_{total}} \label{eq21a} \\
		\text { st. } &R_{g}^m \geq R_{t h}^m, g \in\mathcal{G}, \label{eq21b}\\
		&R_{k}^u \geq R_{t h}^u, k \in\mathcal{K}, \label{eq21c}\\
		&\left|\phi_n\right|=1, n\in\mathcal{N}, \label{eq21d}\\
		&\Vert\boldsymbol{w}_c\Vert^2+\sum_{g\in\mathcal{G}}\Vert\boldsymbol{w}_{p,g}^m\Vert^2+\sum_{k \in\mathcal{K}}\Vert\boldsymbol{w}_{p,k}^u\Vert^2 \leq P_{\text {max }}, \label{eq21e}\\
		&R_{c,k}\geq \sum_{k\in\mathcal{K}} C_{c,k}+\sum_{ g\in\mathcal{G}} C_{c,g}, \label{eq21f}\\
		& C_{c,g} \geq 0, g \in\mathcal{G}, C_{c,k} \geq 0, k \in\mathcal{K},\label{eq21g}
	\end{align}
\end{subequations}
where $\boldsymbol{C}=[C_{c,1}^m,\cdots,C_{c,G}^m,C_{c,1}^u,\cdots,C_{c,K}^u]$ is the vector of the common rate portions. Constraints \eqref{eq21b} and \eqref{eq21c} impose the QoS requirements for unicast and multicast services, respectively. \eqref{eq21d} is the unit-modulus constraint for the RIS elements. \eqref{eq21e} ensures that the maximum transmit power is not exceeded. \eqref{eq21f} guarantees that the common stream $s_c$ can be successfully decoded by all users, while \eqref{eq21g} ensures the non-negativity of the common rate portions allocated to both multicast and unicast messages. Note that the formulated problem is nonconvex due to the nonconvex objective function and constraints. 

\subsubsection{Transformation of Objective Function}First, we introduce an auxiliary nonnegative variable $\varepsilon$ to convert the fractional objective into the polynomial expression by employing the Dinkelbach algorithm\cite{intro30}, then $\mathcal{P}_1$ can be transformed to
\begin{subequations}
	\begin{align}
		\mathcal{P}_2:\max _{\{\boldsymbol{W},\boldsymbol{\phi},\boldsymbol{C},\varepsilon\}} &\sum\nolimits_{k\in\mathcal{K}} R_{k}^u+\sum\nolimits_{g\in\mathcal{G}} R_g^m-\varepsilon P_{total}\label{eq22a} \\
		\text { st. } &\eqref{eq21b}-\eqref{eq21g}, \label{eq22b}
	\end{align}
\end{subequations}
where the optimal system EE achievable under the current resource allocation $\varepsilon^{\star}$ is updated by
\begin{equation}\label{eq23}
	\varepsilon^{\star} = \frac{\sum\limits_{k\in\mathcal{K}} ((C_{c,k}^{u})^{\star}+(R_{p,k}^{u})^{\star})+\sum\limits_{g\in\mathcal{G}} ((C_{c,g}^{m})^{\star}+(R_{p,g}^{m})^{\star})}{\eta P_{total}^{\star}}.
\end{equation}
where $(\cdot)^{\star}$ denote the optimal solution of the associated optimization problem. In each iteration, $\varepsilon$ is updated by maximizing the subtractive objective function given in \eqref{eq22a}. Upon convergence to a local optimum, this subtractive objective function reaches zero. At this point, $\varepsilon^*$ corresponds to the maximum attainable EE, and the associated resource allocation simultaneously optimizes the original EE maximization problem. Next, we address the subproblem $\mathcal{P}_2$ to update the original variables $\{\boldsymbol{W},\boldsymbol{C},\boldsymbol{\Phi}\}$.

\subsubsection{Update $\{\boldsymbol{W},\boldsymbol{C},\boldsymbol{\Phi}\}$}
To start with, we introduce new slack variables $\boldsymbol{\gamma}_{p}^m=[\gamma_{p,1}^m,\cdots,\gamma_{p,G}^m]$, $\boldsymbol{\gamma}_{p}^u=[\gamma_{p,1}^u,\cdots,\gamma_{p,K}^u]$ and $\gamma_c$, to deal with the nonconvex rate constraints, such as \eqref{eq21b}, \eqref{eq21c}, \eqref{eq21f}, \eqref{eq22a}, then $\mathcal{P}_2$ is reduced to 
\begin{subequations}
	\begin{align}
		\mathcal{P}_3:&\max_{\substack{\{
				\boldsymbol{W},\boldsymbol{C},\\ 
				\boldsymbol{\gamma}_{g}^m,\boldsymbol{\gamma}_{p}^u,\gamma_c\}}} \sum\nolimits_{k\in\mathcal{K}} \left(C_{u,k}^u+\log_2(1+\gamma_{p,k}^u)\right)\notag\\
		&+\sum\nolimits_{g\in\mathcal{G}}\left(C_{c,g}^m+\log_2(1+\gamma_{p,g}^m)\right)-\varepsilon P_{total} \\
		\text { st. } &\eqref{eq21d},\eqref{eq21e},\eqref{eq21g},\label{eq24b}\\
		&C_{c,g}^m+\log_2(1+\gamma_{p,g}^m) \geq R_{t h}^m, g\in\mathcal{G},\label{eq24c} \\
		&C_{u,k}^u+\log_2(1+\gamma_{p,k}^u) \geq R_{t h}^u, k\in\mathcal{K},\label{eq24d} \\
		&\log_2(1+\gamma_{c})\geq \sum\nolimits_{k\in\mathcal{K}} C_{c,k}^u+\sum\nolimits_{ g\in\mathcal{G}} C_{c,g}^m,\label{eq24e}\\
		&\frac{\left|(\boldsymbol{\phi}^H\boldsymbol{F}_k+\boldsymbol{d}_k^H)\boldsymbol{w}_{p,g}^m\right|^2}{I_{p,k}^{m}+\sigma_k^2}\geq \gamma_{p,g}^{m},k\in\mathcal{K}_g, g\in\mathcal{G},\label{eq24f}\\
		&\frac{\left|(\boldsymbol{\phi}^H\boldsymbol{F}_k+\boldsymbol{d}_k^H)\boldsymbol{w}_{p,k}^u\right|^2}{I_{p,k}^u+\sigma_k^2}\geq \gamma_{p,k}^u,k\in\mathcal{K},\label{eq24g}\\
		&\frac{\left|(\boldsymbol{\phi}^H\boldsymbol{F}_k+\boldsymbol{d}_k^H)\boldsymbol{w}_c\right|^2}{I_{c,k}+\sigma_k^2}\geq \gamma_{c},k\in\mathcal{K}.\label{eq24h}
	\end{align}
\end{subequations}

However, constraints \eqref{eq24f}-\eqref{eq24h} remain nonconvex, making $\mathcal{P}_3$ challenging to solve. Notably, these constraints share a common signal to interference plus noise ratio (SINR) fractional structure. To address this, we propose a convexification approach for one constraint and extend it to the others.

Firstly, we solve the non-convex constraint \eqref{eq24f}. By introducing a slack non-negative variables $\boldsymbol{\tau}_p^m=[\boldsymbol{\tau}_{p,1}^m,\cdots,\boldsymbol{\tau}_{p,G}^m]$, \eqref{eq24f} can be deduced as
\begin{subequations}
	\begin{align}
		&\left|(\boldsymbol{\phi}^H\boldsymbol{F}_k+\boldsymbol{d}_k^H)\boldsymbol{w}_{p,g}^m\right|^2\geq \gamma_{p,g}^m \tau_{p,g}^m,\forall k\in\mathcal{K}_g,g\in\mathcal{G},\label{eq25a}\\
		&\sum_{g'\neq g}\left|(\boldsymbol{\phi}^H\boldsymbol{F}_k+\boldsymbol{d}_k^H)\boldsymbol{w}_{p, g'}^m\right|^2+\sum_{j\in\mathcal{K}}\left|(\boldsymbol{\phi}^H\boldsymbol{F}_k+\boldsymbol{d}_k^H)\boldsymbol{w}_{p,j}^u\right|^2\notag\\
		&+\sigma_k^2 \leq \tau_{p,g}^m,\forall k\in\mathcal{K}_g,g\in\mathcal{G}.\label{eq25b}
	\end{align}
\end{subequations}
\eqref{eq25a} is still non-convex due to the quadratic form and coupled second term. Taking Taylor's first-order expansion at the $t$-th iteration point $\boldsymbol{w}_{p,g}^{m,(t)}$, the lower bound of left side of \eqref{eq25a} can be  approximately written as
\begin{equation}\label{eq26}
	\begin{aligned}
		&\left|(\boldsymbol{\phi}^H\boldsymbol{F}_k+\boldsymbol{d}_k^H)\boldsymbol{w}_{p,g}^m\right|^2\\
		\geq& 2\Re\{{(\boldsymbol{w}_{p,g}^{m,(t)}})^H((\boldsymbol{\phi}^{(t)})^H\boldsymbol{F}_k+\boldsymbol{d}_k^H)^H(\boldsymbol{\phi}^H\boldsymbol{F}_k+\boldsymbol{d}_k^H)\boldsymbol{w}_{p,g}^m\}\\
		&-\left|((\boldsymbol{\phi}^{(t)})^H\boldsymbol{F}_k+\boldsymbol{d}_k^H)\boldsymbol{w}_{p,g}^{m,(t)}\right|^2.
	\end{aligned}
\end{equation}
For the right side in \eqref{eq25a}, its upper bound can be approximately given by
\begin{equation}\label{eq27}
	\begin{aligned}
		\gamma_{p,g}^m\tau_{p,g}^m\leq\frac{\gamma_{p,g}^{m,(t)}}{2\tau_{p,g}^{m,(t)}}(\tau_{p,g}^{m})^2+\frac{\tau_{p,g}^{m,(t)}}{2\gamma_{p,g}^{m,(t)}}(\gamma_{p,g}^{m})^2.
	\end{aligned}
\end{equation}
In summary, \eqref{eq24f} is convexified as follows
\begin{equation}\label{eq28}
	\begin{aligned}
		&2\Re\{(\boldsymbol{w}_{p,g}^{m,(t)})^H((\boldsymbol{\phi}^{(t)})^H\boldsymbol{F}_k+\boldsymbol{d}_k^H)^H(\boldsymbol{\phi}^H\boldsymbol{F}_k+\boldsymbol{d}_k^H)\boldsymbol{w}_{p,g}^m\}\\
		&\quad-\left|(\boldsymbol{\phi}^{(t)})^H\boldsymbol{F}_k+\boldsymbol{d}_k^H)\boldsymbol{w}_{p,g}^{m,(t)}\right|^2\\
		&\geq\frac{\gamma_{p,g}^{m,(t)}}{2\tau_{p,g}^{m,(t)}}(\tau_{p,g}^{m})^2+\frac{\tau_{p,g}^{m,(t)}}{2\gamma_{p,g}^{m,(t)}}(\gamma_{p,g}^{m})^2,k\in\mathcal{K}_g,g\in\mathcal{G}.
	\end{aligned}
\end{equation}

Similarly, by introducing new non-negative slack variables $\boldsymbol{\tau}_p^u=[\boldsymbol{\tau}_{p,1}^u,\cdots,\boldsymbol{\tau}_{p,K}^u]$ and $\boldsymbol{\tau}_c=[\boldsymbol{\tau}_{c,1},\cdots,\boldsymbol{\tau}_{c,K}]$, \eqref{eq24g}-\eqref{eq24h} are approximately convexified as follows, respectively
\begin{subequations}
	\begin{align}
		&2\Re\{(\boldsymbol{w}_{p,k}^{u,(t)})^H((\boldsymbol{\phi}^{(t)})^H\boldsymbol{F}_k+\boldsymbol{d}_k^H)^H(\boldsymbol{\phi}^H\boldsymbol{F}_k+\boldsymbol{d}_k^H)\boldsymbol{w}_{p,k}^{u}\}\notag\\
		&-\left|((\boldsymbol{\phi}^{(t)})^H\boldsymbol{F}_k+\boldsymbol{d}_k^H)\boldsymbol{w}_{p,k}^{u,(t)}\right|^2\geq\frac{\gamma_{p,k}^{u,(t)}}{2\tau_{p,k}^{u,(t)}}(\tau_{p,k}^{u})^2+\frac{\tau_{p,k}^{u,(t)}}{2\gamma_{p,k}^{u,(t)}}(\gamma_{p,k}^{u})^2,\label{eq29a}\\
		&\sum\limits_{g\neq g_k}\left|(\boldsymbol{\phi}^H\boldsymbol{F}_k+\boldsymbol{d}_k^H)\boldsymbol{w}_{p, g}^m\right|^2+\sum\limits_{j\neq k}\left|(\boldsymbol{\phi}^H\boldsymbol{F}_k+\boldsymbol{d}_k^H)\boldsymbol{w}_{p,j}^u\right|^2\notag \\
		&+\sigma_k^2\leq \tau_{p,k}^u,k\in\mathcal{K},\label{eq29b}\\
		&2\Re\{(\boldsymbol{w}_c^{(t)})^H((\boldsymbol{\phi}^{(t)})^H\boldsymbol{F}_k+\boldsymbol{d}_k^H)^H(\boldsymbol{\phi}^H\boldsymbol{F}_k+\boldsymbol{d}_k^H)\boldsymbol{w}_{c}\}\notag\\
		&-\left|(\boldsymbol{\phi}^{(t)})^H\boldsymbol{F}_k+\boldsymbol{d}_k^H)\boldsymbol{w}_{c}^{(t)}\right|^2\geq\frac{\gamma_{c}^{(t)}}{2\tau_{c,k}^{(t)}}(\tau_{c,k})^2+\frac{\tau_{c,k}^{(t)}}{2\gamma_{c}^{(t)}}(\gamma_{c})^2,\label{eq29c}\\
		&\sum_{g\in\mathcal{G}}\left|(\boldsymbol{\phi}^H\boldsymbol{F}_k+\boldsymbol{d}_k^H) \boldsymbol{w}_{p,g}^m\right|^2+\sum_{j\in\mathcal{K}}\left|(\boldsymbol{\phi}^H\boldsymbol{F}_k+\boldsymbol{d}_k^H) \boldsymbol{w}_{p,j}^u\right|^2\notag \\
		&+\sigma_k^2\leq \tau_{c,k},k\in\mathcal{K}\label{eq29d}.
	\end{align}
\end{subequations}

Based on the preceding discussions, $\mathcal{P}_2$ is reformulated as the approximate problem
\begin{equation}
	\begin{aligned}
		\widetilde{\mathcal{P}}_3:\max_{\substack{\{
				\boldsymbol{W},\boldsymbol{C},\boldsymbol{\phi},\boldsymbol{\gamma}_{g}^m,\\
				\boldsymbol{\gamma}_{p}^u,\boldsymbol{\gamma}_c,
				\boldsymbol{\tau}_{g}^m,\boldsymbol{\tau}_{p}^u,\boldsymbol{\tau}_c\}}} &\sum\nolimits_{k\in\mathcal{K}} \left(C_{u,k}^u+\log_2(1+\gamma_{p,k}^u)\right)\notag\\
		&+\sum\nolimits_{g\in\mathcal{G}}\left(C_{c,g}^m+\log_2(1+\gamma_{p,g}^m)\right)-\varepsilon P_{total} \\
	\text { st. } &\eqref{eq21d},\eqref{eq21e},\eqref{eq21g},\eqref{eq24c}-\eqref{eq24e},\\
	&\eqref{eq25b},\eqref{eq28},\eqref{eq29a}-\eqref{eq29d}. 
	\end{aligned}
\end{equation}
Due to the coupling between $\boldsymbol{W}$ and $\boldsymbol{\phi}$ in constraints \eqref{eq25b},\eqref{eq28},\eqref{eq29a}-\eqref{eq29d}, $\widetilde{\mathcal{P}}_3$ remains nonconvex. To tackle $\widetilde{\mathcal{P}}_3$, We therefore adopt BCA approach, iteratively optimizing the variables $\{\boldsymbol{W}, \boldsymbol{C}\}$ and $\boldsymbol{\phi}$. Specifically, with given $\boldsymbol{\phi}$, the subproblem of $\{\boldsymbol{W}, \boldsymbol{C}\}$ is 
\begin{equation}
	\begin{aligned}
		\mathcal{P}_4:\max_{\substack{\{
				\boldsymbol{W},\boldsymbol{C},\boldsymbol{\gamma}_{g}^m,\\
				\boldsymbol{\gamma}_{p}^u,\gamma_c,\boldsymbol{\tau}_{g}^m,
				\boldsymbol{\tau}_{p}^u,\boldsymbol{\tau}_c\}}} &\sum\nolimits_{k\in\mathcal{K}} \left(C_{u,k}^u+\log_2(1+\gamma_{p,k}^u)\right)\notag\\
		&+\sum\nolimits_{g\in\mathcal{G}}\left(C_{c,g}^m+\log_2(1+\gamma_{p,g}^m)\right)-\varepsilon P_{total} \\
		\text { st. } &\eqref{eq21e},\eqref{eq21g},\eqref{eq24c}-\eqref{eq24e},\eqref{eq25b},\eqref{eq28},\\
		&\eqref{eq29a}-\eqref{eq29d}. 
	\end{aligned}
\end{equation}
So far, $\mathcal{P}_4$ is convex and can be solve by the CVX tools.

In the subproblem of $\boldsymbol{\phi}$, the unit module constraint \eqref{eq21d} is non-convex. We reformulate $\widetilde{\mathcal{P}}_3$ via the penalty method \cite{intro30_penalty} as
\begin{equation}\notag
	\begin{aligned}
		\widetilde{\mathcal{P}}_4:
		\max_{\substack{\{
				\boldsymbol{\phi},\boldsymbol{\gamma}_{g}^m,\boldsymbol{\gamma}_{p}^u,\gamma_c,\\
				\boldsymbol{\tau}_{g}^m,\boldsymbol{\tau}_{p}^u,\boldsymbol{\tau}_c\}}} &\sum\nolimits_{k\in\mathcal{K}} \left(C_{u,k}^u+\log_2(1+\delta_{p,k}^u)\right)\\
		&+\sum\nolimits_{g\in\mathcal{G}}\left(C_{c,g}^m+\log_2(1+\delta_{p,g}^m)\right)\\
		&+\mu\sum_{n\in\mathcal{N}}(2\Re\{(\phi_n^{(t)})^*\phi_n\}-\vert\phi_n^{(t)}\vert^2-1)\\
		\text { st. } &\eqref{eq24c}-\eqref{eq24e},\eqref{eq25b},\eqref{eq28},\eqref{eq29a}-\eqref{eq29d},\\
		&\left|\phi_n\right|^2\leq 1, \forall n,
	\end{aligned}
\end{equation}
where $\mu$ is a positive penalty factor. $\widetilde{\mathcal{P}}_4$ is convex and can be effectively solved via CVX with MOSEK. 

In each iteration, the active and passive beamforming vectors and EE value are updated alternately until a convergence criterion is met.
The overall SCA-based BCA algorithm for solving $\mathcal{P}_1$ is summarized in \textbf{Algorithm \ref{Alg1}}.
\begin{algorithm}[htb]
	\caption{Proposed SCA-based BCA algorithm.}
	\label{Alg1}
	\begin{algorithmic}[1]
		\STATE Initialize variables $\{\boldsymbol{W}^{(0)},\boldsymbol{\phi}^{(0)},\boldsymbol{\gamma}_{p}^{m,(0)},
		\boldsymbol{\gamma}_{p}^{u,(0)},\gamma_c^{(0)},\boldsymbol{\tau}_{p}^{m,(0)},$
		$\boldsymbol{\tau}_{p}^{u,(0)},\tau_c^{(0)}\}$, $\varepsilon^{(0)}=0$, penalty factor $\mu$, iteration number $t=0$. 
		\REPEAT
		\STATE For given $\{\boldsymbol{\phi}^{(t)},\varepsilon^{(t)}\}$, update the optimal $\{\boldsymbol{W}^{(t+1)},\boldsymbol{C}^{(t+1)}\}$ by solving $\mathcal{P}_4$.
		\STATE For given $\{\boldsymbol{W}^{(t+1)},\boldsymbol{C}^{(t+1)},\varepsilon^{(t)}\}$, update the optimal $\{\boldsymbol{\phi}^{(t+1)}\}$ by solving $\widetilde{\mathcal{P}}_4$.
		\STATE Update $,\varepsilon^{(t+1)}$ based on \eqref{eq23}.
		\STATE $t=t+1$.
		\UNTIL{ the objective function of $\mathcal{P}_1$ is convergent. }
		\STATE Output $\{\boldsymbol{W}^{(t+1)},\boldsymbol{C}^{(t+1)},\boldsymbol{\phi}^{(t)}\}$.  
	\end{algorithmic}
\end{algorithm}
\subsection{Joint Beamforming Design for PCF-RSMA}
In this scheme, the optimization problem of EE maximization is written as 
\begin{subequations}
	\begin{align}
		\mathcal{P}_5:
		\max _{\{\acute{\boldsymbol{W}}, \acute{\boldsymbol{\phi}},\acute{\boldsymbol{C}}\}} &\frac{\sum\nolimits_{k\in\mathcal{K}} \acute{R}_{k}^u+\sum\nolimits_{g\in\mathcal{G}} \acute{R}_g^m}{\acute{P}_{total}} \label{eq34a} \\
		\text { st. } &\acute{R}_{g}^m \geq R_{t h}^m, g\in\mathcal{G}, \label{eq32b}\\
		&\acute{R}_{k}^u \geq R_{t h}^u, k\in\mathcal{K}, \label{eq32c}\\
		&\left|\acute{\phi}_n\right|=1, n\in\mathcal{N}, \label{eq32d}\\
		&\Vert\acute{\boldsymbol{w}}_c\Vert^2+\sum_{g\in\mathcal{G}}\Vert\acute{\boldsymbol{w}}_{g}^{gs}\Vert^2+\sum_{k\in\mathcal{K}}\Vert\acute{\boldsymbol{w}}_{p,k}^u\Vert^2 \leq P_{\text {max }}, \label{eq32e}\\
		&\acute{R}_{c,k}\geq\sum\nolimits_{g\in\mathcal{G}}\acute{C}_{c,g}^m, k\in\mathcal{K},\label{eq32f}\\
		&\acute{R}_{c,k}^{gs}\geq\sum\nolimits_{k\in \mathcal{K}_g}\acute{C}_{c,k}^{gs,u}+\acute{C}_{c,g}^{gs,m}, \label{eq32g}\\
		& \acute{C}_{c,g}^m \geq 0, g\in\mathcal{G}, \label{eq32h}\\
		&\acute{C}_{c,k}^{gs,u} \geq 0, k\in\mathcal{K}_g,g\in\mathcal{G},\acute{C}_{c,g}^{gs,m}\geq 0, g\in\mathcal{G}.\label{eq32i}
	\end{align}
\end{subequations}
where $\acute{\boldsymbol{C}}\triangleq\{\acute{C}_{c,1}^m,\cdots,\acute{C}_{c,G}^m,\acute{C}_{c,1}^u,\cdots,\acute{C}_{c,K}^u,C_{c,1}^{gs,m},\cdots,\\
C_{c,G}^{gs,m},\underbrace{\acute{C}_{c,1}^{gs,u},\cdots,\acute{C}_{c,|\mathcal{K}_g|}^{gs,u}}_{g\in\mathcal{G}}\}$ is the vector of all common rate portions across the global and group-specific common streams. Constraints \eqref{eq32b} and \eqref{eq32c} impose unicast and multicast QoS demands. \eqref{eq32d} enforces unit-modulus on RIS elements. Inequality \eqref{eq32e} limits the total transmit power. Constraints \eqref{eq32f} and \eqref{eq32g} guarantee that the common stream $\acute{s}_c$ and the group-specific common stream $\acute{s}_{c,g}^{gs},g\in\mathcal{G}$ can be successfully decoded by all users and the corresponding group users, respectively. Finally, \eqref{eq32h} and \eqref{eq32i} enforce the non-negativity of the common rate portions allocated to the global and group-specific messages for unicast and multicast services.

The formulated EE optimization problem $\mathcal{P}_5$ for PCF-RSMA is also non-convex. It shares the same objective function and constraint structure as $\mathcal{P}_1$, the key difference lies in the underlying signal model. Specifically, $\mathcal{P}_1$ involves a single global common stream, whereas $\mathcal{P}_5$ features a global and a group-level common stream. However, the SCA-based BCA algorithm developed for $\mathcal{P}_1$ remains applicable with only minor modifications to the stream-specific rate expressions. We thus omit the repetitive algorithmic details for brevity.
\vspace{-12pt}
\section{Robust JOINT BEAMFORMING DESIGN}
This section investigates joint beamforming design under practical non-ideal CSI for both CCF‑RSMA and PCF‑RSMA schemes. Although various channel estimation methods\cite{intro31,intro32,intro33} have been proposed for RIS‑assisted mmWave systems, obtaining accurate CSI remains challenging due to the passive nature of RIS and user mobility. Therefore, we consider robust beamforming designs that account for CSI uncertainties.

We thus adopt a bounded CSI error model to characterize channel uncertainty through an error region of specified radius, ensuring worst‑case performance guarantees. Based on practical deployment characteristics, we assume perfect CSI for the BS-RIS link since both ends are typically fixed. In contrast, due to user mobility, the direct BS-user and RIS-user links are subject to imperfect CSI. Accordingly, the direct channel $\boldsymbol{d}_k$ and the RIS-user channel $\boldsymbol{h}_k$ can be respectively expressed as
\begin{equation}\label{eq33}
	\begin{aligned}
		&\boldsymbol{d}_k=\hat{\boldsymbol{d}}_k+\Delta \boldsymbol{d}_k,\Delta \boldsymbol{d}_k ,\left\|\Delta \boldsymbol{d}_k\right\|_2 \leq \xi_{\boldsymbol{d}_k},k\in\mathcal{K},\\
		&\boldsymbol{h}_k  =\hat{\boldsymbol{h}}_k+\Delta \boldsymbol{h}_k,\Delta \boldsymbol{h}_k ,\left\|\Delta \boldsymbol{h}_k\right\|_2 \leq \xi_{\boldsymbol{h}_k} ,k\in\mathcal{K},
	\end{aligned}
\end{equation}
where $\hat{\boldsymbol{d}}_k$ and $\hat{\boldsymbol{h}}_k$ are the direct and partial cascaded estimated CSI of user $k$ known at the BS, respectively. $\Delta \boldsymbol{d}_k$ and $\Delta \boldsymbol{h}_k$ are the corresponding unknown CSI estimation errors, $\xi_{\boldsymbol{d}_k}$ and $\xi_{\boldsymbol{h}_k}$ are the uncertainty region known at the BS. 

Under this model, the EE maximization problem for CCF-RSMA scheme is the same as $\mathcal{P}_1$. Denote $\zeta_k\triangleq\{\left\|\Delta \boldsymbol{d}_k\right\|_2 \leq \xi_{\boldsymbol{d}_k},\left\|\Delta \boldsymbol{h}_k\right\|_2 \leq \xi_{\boldsymbol{h}_k},k\in\mathcal{K}\}$. By introducing non-negative auxiliary variables $\varepsilon$ for EE value need to update during per iteration, $\boldsymbol{\gamma}_p^m=[\gamma_{p,1}^m,\cdots,\gamma_{p,G}^m]$, $\boldsymbol{\gamma}_p^u=[\gamma_{p,1}^u,\cdots,\gamma_{p,K}^u]$, $\gamma_c$ for SINR, the non-negative auxiliary variables $\boldsymbol{\alpha}_p^m=[\alpha_{p,1}^m,\cdots,\alpha_{p,G}^m]$, $\boldsymbol{\alpha}_p^u=[\alpha_{p,1}^u,\cdots,\alpha_{p,K}^u]$, $\boldsymbol{\alpha}_p^m=[\alpha_{c,1},\cdots,\alpha_{c,K}]$ and $\boldsymbol{\beta}_p^m=[\beta_{p,1}^m,\cdots, \beta_{p,G}^m]$, $\boldsymbol{\beta}_p^m=[\beta_{p,1}^u,\cdots, \beta_{p,K}^u]$, $\boldsymbol{\beta}_p^m=[\beta_{c,1},\cdots, \beta_{c,K}]$ for the fractional forms of the SINR expressions, the original $\mathcal{P}_1$ can be equicalently transformed into $\mathcal{P}_6$ at the top of this page. 
\begin{figure*}[!t]
	\normalsize
	\setcounter{equation}{31}
	\begin{subequations}
		\begin{align}
      \mathcal{P}_6:\max_{\substack{\{
		\boldsymbol{W},\boldsymbol{C},\boldsymbol{\Phi},\varepsilon,\boldsymbol{\gamma}_{p}^m,\boldsymbol{\gamma}_{p}^u,\gamma_c, \\
		\boldsymbol{\alpha}_{p}^m,\boldsymbol{\alpha}_{p}^u,\boldsymbol{\alpha}_c,\boldsymbol{\beta}_{p}^m,\boldsymbol{\beta}_{p}^u,\boldsymbol{\beta}_c\}}} &\sum\nolimits_{k\in\mathcal{K}} \left(C_{u,k}^u+\log_2(1+\gamma_{p,k}^u)\right)+\sum\nolimits_{g\in\mathcal{G}}\left(C_{c,g}^m+\log_2(1+\gamma_{p,g}^m)\right)-\varepsilon P_{total} \\
        \text { st. } &\eqref{eq21d},\eqref{eq21e},\eqref{eq21g},\label{eq34b}\\
        &C_{c,g}^m+\log_2(1+\gamma_{p,g}^m) \geq R_{t h}^m, g\in\mathcal{G},\label{eq34c} \\
        &C_{u,k}+\log_2(1+\gamma_{p,k}^u) \geq R_{t h}^u, k\in\mathcal{K}, \label{eq34d}\\
        &\log_2(1+\gamma_{c})\geq \sum\nolimits_{k\in\mathcal{K}} C_{c,k}+\sum\nolimits_{ g\in\mathcal{G}} C_{c,g},\label{eq34e}\\
        &\left|(\boldsymbol{h}_k^H\boldsymbol{\Phi}\boldsymbol{H}+\boldsymbol{d}_k^H)\boldsymbol{w}_{p,g}^m\right|^2\geq\alpha_{p,g}^m,\zeta_k,k\in\mathcal{K}_g, g\in\mathcal{G},\label{eq34f}\\
        &\sum\limits_{g'\neq g}\left|(\boldsymbol{h}_k^H\boldsymbol{\Phi}\boldsymbol{H}+\boldsymbol{d}_k^H)\boldsymbol{w}_{p, g'}^m\right|^2+\sum\limits_{j\in\mathcal{K}}\left|(\boldsymbol{h}_k^H\boldsymbol{\Phi}\boldsymbol{H}+\boldsymbol{d}_k^H)\boldsymbol{w}_{p,j}^u\right|^2+\sigma_k^2\leq \beta_{p,g}^m,\zeta_k,k\in\mathcal{K}_g, g\in\mathcal{G},\label{eq34g}\\
        &\gamma_{p,g}^m\beta_{p,g}^m\leq\alpha_{p,g}^m, g\in\mathcal{G},\label{eq34h}\\
        &\left|(\boldsymbol{h}_k^H\boldsymbol{\Phi}\boldsymbol{H}+\boldsymbol{d}_k^H)\boldsymbol{w}_{p,k}^u\right|^2\geq \alpha_{p,k}^u,\zeta_k,k\in\mathcal{K},\label{eq34i}\\
        &\sum\nolimits_{g\neq g_k}\left|(\boldsymbol{h}_k^H\boldsymbol{\Phi}\boldsymbol{H}+\boldsymbol{d}_k^H)\boldsymbol{w}_{p, g}^m\right|^2+\sum\nolimits_{j\neq k}\left|(\boldsymbol{h}_k^H\boldsymbol{\Phi}\boldsymbol{H}+\boldsymbol{d}_k^H)\boldsymbol{w}_{p,j}^u\right|^2+\sigma_k^2\leq \beta_{p,k}^u,\zeta_k,k\in\mathcal{K},\label{eq34j}\\
        &\gamma_{p,k}^u\beta_{p,k}^u\leq\alpha_{p,k}^u,k\in\mathcal{K},\label{eq34k}\\
        &\left|(\boldsymbol{h}_k^H\boldsymbol{\Phi}\boldsymbol{H}+\boldsymbol{d}_k^H)\boldsymbol{w}_c\right|^2\geq \alpha_{c,k},\zeta_k,k\in\mathcal{K},\label{eq34l}\\
        &\sum\nolimits_{g\in\mathcal{G}}\left|(\boldsymbol{h}_k^H\boldsymbol{\Phi}\boldsymbol{H}+\boldsymbol{d}_k^H) \boldsymbol{w}_{p,g}^m\right|^2+\sum\nolimits_{j\in\mathcal{K}}\left|(\boldsymbol{h}_k^H\boldsymbol{\Phi}\boldsymbol{H}+\boldsymbol{d}_k^H) \boldsymbol{w}_{p,j}^u\right|^2+\sigma_k^2\leq  \beta_{c,k},\zeta_k,k\in\mathcal{K},\label{eq34m}\\
        &\gamma_{c}\beta_{c,k}\leq\alpha_{c,k},k\in\mathcal{K}\label{eq34n}.
\end{align}
\end{subequations}
\hrulefill
\vspace*{0pt}
\end{figure*}

$\mathcal{P}_6$ involves semi-infinite constraints due to channel uncertainties, and non-convex coupling between active and passive beamforming. We employ a BCA manner. With RIS phase shifts fixed, the S-Procedure converts robust constraints into LMIs. With beamforming and rate allocation fixed, the S-Procedure again handles robust constraints, while the penalty method enforces unit-modulus constraints. The algorithm iterates until convergence to a stationary point.

Before designing beamforming, we first elaborate a general Lemma given as follows.
\begin{lemma}[General S-Procedure, Lemma 1 in \cite{alg1}]
	Let $\boldsymbol{A}\in\mathbb{H}^{n\times n}, \boldsymbol{a}\in\mathbb{C}^{n\times 1}, a\in\mathbb{R}, \boldsymbol{z}\in\mathbb{C}^{n\times 1}$, then a quadratic function $f_i(\boldsymbol{z})$ is defined as
	\begin{equation}
		f_i(\boldsymbol{z}) = \boldsymbol{z}^H \boldsymbol{A}_i \boldsymbol{z} + 2\operatorname{Re}\{\boldsymbol{a}_i^H \boldsymbol{z}\} + a_i, \quad i = \{0,1\},
	\end{equation}
	The condition $f_0(\boldsymbol{z}) \geq 0 \Rightarrow f_1(\boldsymbol{z}) \geq 0$ holds if and only if there exist $\varpi \geq 0$ such that
	\begin{equation}
		\begin{bmatrix}
			\boldsymbol{A}_0 & \boldsymbol{a}_0 \\
			\boldsymbol{a}_0^H & a_0
		\end{bmatrix}
		-  
		\varpi\begin{bmatrix}
			\boldsymbol{A}_1 & \boldsymbol{a}_1 \\
			\boldsymbol{a}_1^H & a_1
		\end{bmatrix} \succeq \boldsymbol{0}.
	\end{equation}
\end{lemma}
\vspace{-15pt}
\subsection{Active beamforming design}
With the RIS phase-shift matrix fixed, we first address the
active beamforming optimization.  The constraints \eqref{eq34f}-\eqref{eq34n} exhibit a repeating pattern across multicast private, unicast private, and common streams. Therefore, we first detail the convexification for the multicast SINR constraints \eqref{eq34f}-\eqref{eq34h}, then the remaining constraints are handled analogously.

We define the rank-one positive semidefinite matrices $\boldsymbol{W}_{p,g}^m=\boldsymbol{w}_{p,g}^m(\boldsymbol{w}_{p,g}^{m})^H$, $\boldsymbol{W}_{p,k}^u=\boldsymbol{w}_{p,k}^u(\boldsymbol{w}_{p,k}^{u})^H$ and $\boldsymbol{W}_{c}=\boldsymbol{w}_{c}\boldsymbol{w}_{c}^{H}$. Substituting \eqref{eq33} into \eqref{eq34f} yields
\begin{equation}\label{eq37}
	\begin{aligned}
	&\Delta\boldsymbol{x}_k^H\tilde{\boldsymbol{W}}_{p,g}^m\Delta\boldsymbol{x}_k+2\Re\{\hat{\boldsymbol{x}}_k^H\tilde{\boldsymbol{W}}_{p,g}^m\Delta\boldsymbol{x}_k\}+\hat{\boldsymbol{x}}_k^H\tilde{\boldsymbol{W}}_{p,g}^m\hat{\boldsymbol{x}}_k\geq\alpha_{p,g}^m,\\
	&\left\|\Delta \boldsymbol{x}_k\right\|_2 ^2\leq \xi_{\boldsymbol{h}_k}^2+ \xi_{\boldsymbol{d}_k}^2,k\in\mathcal{K}_g, g\in\mathcal{G},
	\end{aligned}
\end{equation}
where $\Delta\boldsymbol{x}_k=[\Delta\boldsymbol{h}_k;\Delta\boldsymbol{d}_k]$, $\hat{\boldsymbol{x}}_k=[\hat{\boldsymbol{h}}_k;\hat{\boldsymbol{d}}_k]$, and $\tilde{\boldsymbol{W}}_{p,g}^m=[\boldsymbol{\Phi}\boldsymbol{H};\boldsymbol{I}_{M}]\boldsymbol{W}_{p,g}^m[\boldsymbol{\Phi}\boldsymbol{H};\boldsymbol{I}_{M}]^H$. Then \eqref{eq37} with CSI uncertainties can be converted into the following LMI based on \textbf{Lemma 1} for $k\in\mathcal{K}_g, g\in\mathcal{G}$,
\begin{equation}\label{eq38}
		\mathcal{L}_{lb}(\tilde{\boldsymbol{W}}_{p,g}^m,\varpi_{p,g}^m,q_{p,g}^m)\triangleq\begin{bmatrix}
			\tilde{\boldsymbol{W}}_{p,g}^m+\varpi_{p,g}^m\boldsymbol{I}_{M+N} & (\tilde{\boldsymbol{W}}_{p,g}^{m})^H\hat{\boldsymbol{x}}_k \\
			\hat{\boldsymbol{x}}_k^H\tilde{\boldsymbol{W}}_{p,g}^m & q_{p,g}^m
		\end{bmatrix}
	\succeq \boldsymbol{0},
\end{equation}
where $q_{p,g}^m=\hat{\boldsymbol{x}}_k^H\tilde{\boldsymbol{W}}_{p,g}^m\hat{\boldsymbol{x}}_k-\alpha_{p,g}^m-\varpi_{p,g}^m(\xi_{\boldsymbol{h}_k}^2+\xi_{\boldsymbol{d}_k}^2)$, and $\boldsymbol{\varpi}_{p}^m=[\varpi_{p,1}^m,\cdots,\varpi_{p,G}^m]$ is the new slack nonnegative variable. The subscript $lb$ indicates that $\mathcal{L}_{lb}(\cdot)$ enforces the lower bound constraint \eqref{eq37} under CSI uncertainties via the S-procedure. This formulation will be reused throughout the paper with different arguments.

For constraint \eqref{eq34g}, we obtain the following LMI constraint for all $k\in\mathcal{K}_g, g\in\mathcal{G}$ by applying \textbf{Lemma 1}
\begin{equation}\label{eq39}
	\begin{aligned}
		& \mathcal{L}_{ub}\left(\tilde{\boldsymbol{W}}_{int, g}^m, \varpi_{int, g}^m, q_{int, g}^m\right) \\
		& \triangleq\left[\begin{array}{cc}
			\varpi_{int, g}^m \boldsymbol{I}_{M+N}-\tilde{\boldsymbol{W}}_{int, g}^m & -(\tilde{\boldsymbol{W}}_{int, g}^{m})^H \hat{\boldsymbol{x}}_k \\
			-\hat{\boldsymbol{x}}_k^H \tilde{\boldsymbol{W}}_{int, g}^m & q_{int, g}^m
		\end{array}\right] \succeq \boldsymbol{0},
	\end{aligned}
\end{equation}
where $q_{int,g}^m=-\hat{\boldsymbol{x}}_k^H\tilde{\boldsymbol{W}}_{int,g}^m\hat{\boldsymbol{x}}_k-\sigma_k^2+\beta_{p,g}^m-\varpi_{int,g}^m(\xi_{\boldsymbol{h}_k}^2+\xi_{\boldsymbol{d}_k}^2)$, and $\boldsymbol{\varpi}_{int}^m=[\varpi_{int,1}^m,\cdots,\varpi_{int,G}^m]\geq 0$ are slack variables. Moreover, $\tilde{\boldsymbol{W}}_{int,g}^m=[\boldsymbol{\Phi}\boldsymbol{H};\boldsymbol{I}_{M}](\sum_{g'\neq g}\boldsymbol{W}_{p,g'}^m+\sum_{j\in\mathcal{K}}\boldsymbol{W}_{p,j}^u)[\boldsymbol{\Phi}\boldsymbol{H};\boldsymbol{I}_{M}]^H$, and the subscript $int$ denotes the aggregate interference at user $k\in\mathcal{K}_g$, comprising both inter-group multicast interference and unicast interference from all users. The subscript $ub$ stands for upper bound, as $\mathcal{L}_{ub}(\cdot)$ enforces an interference power constraint via the S-procedure. 

Similar to \eqref{eq27}, the coupled second term in \eqref{eq34h} is approximately transformed into a convex constraint
\begin{equation}\label{eq40}
	\begin{aligned}
		f(\gamma_{p,g}^{m},\beta_{p,g}^{m},\alpha_{p,g}^{m})\triangleq\frac{\gamma_{p,g}^{m,(t)}}{2\beta_{p,g}^{m,(t)}}(\beta_{p,g}^{m})^2+\frac{\beta_{p,g}^{m,(t)}}{2\gamma_{p,g}^{m,(t)}}(\gamma_{p,g}^{m})^2\leq\alpha_{p,g}^{m}.
	\end{aligned}
\end{equation}
where $f(\cdot)$ is a convex approximation of the coupled term in \eqref{eq34h} via first-order Taylor expansion. Constraints \eqref{eq34k} and \eqref{eq34n} are handled analogously with their respective variables $(\gamma_{p,k}^u, \beta_{p,k}^u, \alpha_{p,k}^u)$ and $(\gamma_{c}, \beta_{c,k}, \alpha_{c,k})$. 

Similarly, the remaining quadratic constraints are transformed as follows:
\begin{itemize}
	\item For \eqref{eq34i}: $\mathcal{L}_{lb}(\tilde{\boldsymbol{W}}_{p,k}^u,\varpi_{p,k}^u,q_{p,k}^u)\succeq\boldsymbol{0}$, where \(q_{p,k}^u = \hat{\boldsymbol{x}}_k^H \tilde{\boldsymbol{W}}_{p,k}^u \hat{\boldsymbol{x}}_k - \alpha_{p,k}^u - \varpi_{p,k}^u (\xi_{\boldsymbol{h}_k}^2 + \xi_{\boldsymbol{d}_k}^2)\), \(\tilde{\boldsymbol{W}}_{p,k}^u = [\boldsymbol{\Phi} \boldsymbol{H}; \boldsymbol{I}_M] \boldsymbol{W}_{p,k}^u [\boldsymbol{\Phi} \boldsymbol{H}; \boldsymbol{I}_M]^H\).
	\item For \eqref{eq34j}: $\mathcal{L}_{ub}(\tilde{\boldsymbol{W}}_{int,k}^u,\varpi_{int,k}^u,q_{int,k}^u)\succeq\boldsymbol{0}$, where \(q_{int,k}^u = -\hat{\boldsymbol{x}}_k^H \tilde{\boldsymbol{W}}_{int,k}^u \hat{\boldsymbol{x}}_k - \sigma_k^2 + \beta_{p,k}^u - \varpi_{int,k}^u (\xi_{\boldsymbol{h}_k}^2 + \xi_{\boldsymbol{d}_k}^2)\), \(\tilde{\boldsymbol{W}}_{int,k}^u = [\boldsymbol{\Phi} \boldsymbol{H}; \boldsymbol{I}_M] \left( \sum_{g\neq g_k} \boldsymbol{W}_{p,g}^m + \sum_{j\neq k} \boldsymbol{W}_{p,j}^u \right) [\boldsymbol{\Phi} \boldsymbol{H}; \boldsymbol{I}_M]^H\).
	\item For \eqref{eq34l}: $\mathcal{L}_{lb}(\tilde{\boldsymbol{W}}_c,\varpi_{c,k},q_{c,k})\succeq\boldsymbol{0}$, where \(q_{c,k} = \hat{\boldsymbol{x}}_k^H \tilde{\boldsymbol{W}}_c \hat{\boldsymbol{x}}_k - \alpha_{c,k} - \varpi_{c,k} (\xi_{\boldsymbol{h}_k}^2 + \xi_{\boldsymbol{d}_k}^2)\), \(\tilde{\boldsymbol{W}}_c = [\boldsymbol{\Phi} \boldsymbol{H}; \boldsymbol{I}_M] \boldsymbol{W}_c [\boldsymbol{\Phi} \boldsymbol{H}; \boldsymbol{I}_M]^H\).
	\item For \eqref{eq34m}: $\mathcal{L}_{ub}(\tilde{\boldsymbol{W}}_{int,c},\varpi_{int,k}^c,q_{int,k}^c)\succeq\boldsymbol{0}$, where \(q_{int,k}^c = -\hat{\boldsymbol{x}}_k^H \tilde{\boldsymbol{W}}_{int,c} \hat{\boldsymbol{x}}_k - \sigma_k^2 + \beta_{c,k} - \varpi_{int,k}^c (\xi_{\boldsymbol{h}_k}^2 + \xi_{\boldsymbol{d}_k}^2)\), \(\tilde{\boldsymbol{W}}_{int,c} = [\boldsymbol{\Phi} \boldsymbol{H}; \boldsymbol{I}_M] \left( \sum_{g\in\mathcal{G}} \boldsymbol{W}_{p,g}^m + \sum_{j\in\mathcal{K}} \boldsymbol{W}_{p,j}^u \right) [\boldsymbol{\Phi} \boldsymbol{H}; \boldsymbol{I}_M]^H\).
\end{itemize}

The rank‑one constraints on the semidefinite matrices $\boldsymbol{W}_c,\{\boldsymbol{W}_{p,g}^m\}_{g=1}^G,\{\boldsymbol{W}_{p,k}^u\}_{k=1}^K$ are handled via the multi-variable SROCR method \cite{ref1}. For any such matrix $\boldsymbol{X}$ in the set, the algorithm iteratively  promotes a rank-one solution through a sequence of convex relaxations. At iteration $t$, the rank-one constraint is approximated as
\begin{equation}\label{eq41}
	\boldsymbol{\lambda}_{\max }\big(\boldsymbol{X}^{(t)}\big)^H \boldsymbol{X} \boldsymbol{\lambda}_{\max }\big(\boldsymbol{X}^{(t)}\big) \geq \nu^{(t)} \operatorname{Tr}\big(\boldsymbol{X}\big),
\end{equation}
where $\boldsymbol{\lambda}_{\max}\big(\boldsymbol{X}^{(t)}\big)$ is the eigenvector corresponding to the largest eigenvalue of $\boldsymbol{X}^{(t)}$, and $\nu^{(t)}= \boldsymbol{\lambda}_{\max }\big(\boldsymbol{X}^{(t)}\big)/\operatorname{Tr}(\boldsymbol{X}^{(t)})\in [0,1]$. This approximation is applied in parallel to all matrices, ensuring progressive rank‑one feasibility.

Finally, the original problem $\mathcal{P}_6$ is reformulated as a convex optimization problem below
\begin{subequations}\notag
	\begin{align}
		\mathcal{P}_7:&\max_{\mathcal{S}_1} \sum\nolimits_{k\in\mathcal{K}} \big(C_{u,k}^u+\log_2(1+\gamma_{p,k}^u)\big)\notag\\
		&+\sum\nolimits_{g\in\mathcal{G}}\big(C_{c,g}^m+\log_2(1+\gamma_{p,g}^m)\big)-\varepsilon P_{total} \\
		\text { st. } &\eqref{eq21g},\eqref{eq34c}-\eqref{eq34e},\eqref{eq38}-\eqref{eq40},\eqref{eq41},\\
		&\mathcal{L}_{lb}(\tilde{\boldsymbol{W}}_{p,k}^u,\varpi_{p,k}^u,q_{p,k}^u)\succeq\boldsymbol{0},\\
		&\mathcal{L}_{ub}(\tilde{\boldsymbol{W}}_{int,k}^u,\varpi_{int,k}^u,q_{int,k}^u)\succeq\boldsymbol{0},\\
		&\mathcal{L}_{lb}(\tilde{\boldsymbol{W}}_c,\varpi_{c,k},q_{c,k})\succeq\boldsymbol{0},\\
		&\mathcal{L}_{ub}(\tilde{\boldsymbol{W}}_{int,c},\varpi_{int,k}^c,q_{int,k}^c)\succeq\boldsymbol{0},\\
		&f(\gamma_{p,k}^{u},\beta_{p,k}^{u},\alpha_{p,k}^{u})\geq0,f(\gamma_{c},\beta_{p,k},\alpha_{p,k})\geq 0,\\
		&\boldsymbol{W}_c\succeq\boldsymbol{0},\{\boldsymbol{W}_{p,g}^m\}_{g=1}^G\succeq\boldsymbol{0},\{\boldsymbol{W}_{p,k}^u\}_{k=1}^K\succeq\boldsymbol{0},\\
		&\operatorname{Tr}(\boldsymbol{W}_c)+\sum\nolimits_{g\in\mathcal{G}}\operatorname{Tr}(\boldsymbol{W}_{p,g}^m)+\sum\nolimits_{k \in\mathcal{K}}\operatorname{Tr}(\boldsymbol{W}_{p,k}^u)\leq P_{max},
	\end{align}
\end{subequations}
where $\mathcal{S}_1\triangleq\{\boldsymbol{W}_c,\{\boldsymbol{W}_{p,g}^m\}_{g=1}^G,\{\boldsymbol{W}_{p,k}^u\}_{k=1}^K,\boldsymbol{C},\boldsymbol{\gamma}_{p}^m,\boldsymbol{\gamma}_{p}^u,\gamma_c,\boldsymbol{\alpha}_{p}^m,\\
\boldsymbol{\alpha}_{p}^u,\boldsymbol{\alpha}_c,\boldsymbol{\beta}_{p}^m,\boldsymbol{\beta}_{p}^u,\boldsymbol{\beta}_c,\boldsymbol{\varpi}_{p}^m,\boldsymbol{\varpi}_{int}^m,\boldsymbol{\varpi}_{p}^u,\boldsymbol{\varpi}_{int}^u,\boldsymbol{\varpi}_{c},\boldsymbol{\varpi}_{int}^c\}$ is the set of all optimization variables. Since $\mathcal{P}_7$ is convex, it can be directly solved using tools such as CVX with MOSEK. The optimal beamforming matrix obtained from the solution can then undergo singular value decomposition to yield the corresponding active beamforming vectors set $\boldsymbol{W}$.

\subsection{Passive beamforming design}
Given $\{\boldsymbol{W},\boldsymbol{C}\}$, we now focus on optimizing $\boldsymbol{\Phi}$. We begin by convexifying constraints \eqref{eq34f} and \eqref{eq34g}, which serve as the basis for handling the structurally similar constraints \eqref{eq34i}, \eqref{eq34j}, \eqref{eq34l} and \eqref{eq34m}. For the representative constraint \eqref{eq34f}, we first apply the SCA method following \eqref{eq26} and then address its CSI uncertainties via the S‑procedure in \textbf{Lemma 1}.

Specifically, the LMI for \eqref{eq34f} is given by
\begin{equation}\label{eq43}
		\mathcal{F}_{lb}(\boldsymbol{\Xi}_{g,k}^m,\varphi_{p,g}^m,\delta_{p,g}^m)\triangleq\begin{bmatrix}
		\boldsymbol{\Xi}_{g,k}^m+\varphi_{p,g}^m\boldsymbol{I}_{M+N} & \boldsymbol{\Xi}_{g,k}^m\hat{\boldsymbol{x}}_k \\
		\hat{\boldsymbol{x}}_k^H\boldsymbol{\Xi}_{g,k}^m & \delta_{p,g}^m
	\end{bmatrix}
	\succeq \boldsymbol{0},
\end{equation}
where $\boldsymbol{\varphi}=[\varphi_{p,1}^m,\cdots,\varphi_{p,G}^m]$ is nonegative slack variable and $\delta_{p,g}^m=\hat{\boldsymbol{x}}_k^H\boldsymbol{\Xi}_{g,p}^m\hat{\boldsymbol{x}}_k-\alpha_{p,g}^m-\varphi_{p,g}^m(\xi_{\boldsymbol{h}_k}^2+\xi_{\boldsymbol{d}_k}^2)$. In particular, $\boldsymbol{\Xi}_{g,p}^m=\boldsymbol{\Psi}_{g,p}^m(\boldsymbol{\Psi}_{g,p}^{m,(t)})^H+\boldsymbol{\Psi}_{g,k}^{m,(t)}(\boldsymbol{\Psi}_{g,p}^{m})^H-\boldsymbol{\Psi}_{g,p}^{m,(t)}(\boldsymbol{\Psi}_{g,p}^{m,(t)})^H$, where $\boldsymbol{\Psi}_{g,p}^m=[\boldsymbol{\Phi}\boldsymbol{H}\boldsymbol{w}_{p,g}^m;\boldsymbol{I}_{M}]$ and $\boldsymbol{\Psi}_{g,p}^{m,(t)}=[\boldsymbol{\Phi}^{(t)}\boldsymbol{H}\boldsymbol{w}_{p,g}^m;\boldsymbol{I}_{M}]$. Similar to $\mathcal{L}_{lb}(\cdot)$ in \eqref{eq38}, the LMI $\mathcal{F}_{lb}(\cdot)$ enforces the same lower bound constraint \eqref{eq34f} but is derived directly from vector-valued variables for RIS phase optimization.

Next, using the same methodology, \eqref{eq34g} is equivalently transformed into the LMI
\begin{equation}\label{eq44}
	\begin{aligned}
		& \mathcal{F}_{ub}\left(\tilde{\boldsymbol{\Xi}}_{int, g}^m, \varphi_{int, g}^m, \delta_{int, g}^m\right) \\
		& \triangleq\left[\begin{array}{cc}
			\varphi_{int, g}^m \boldsymbol{I}_{M+N}-\tilde{\boldsymbol{\Xi}}_{int, g}^m & -\tilde{\boldsymbol{\Xi}}_{int, g}^{m, H} \hat{\boldsymbol{x}}_k \\
			-\hat{\boldsymbol{x}}_k^H \tilde{\boldsymbol{\Xi}}_{int, g}^m & \delta_{int, g}^m
		\end{array}\right] \succeq \boldsymbol{0},
	\end{aligned}
\end{equation}
where $\delta_{int,g}^m=-\hat{\boldsymbol{x}}_k^H\tilde{\boldsymbol{\Xi}}_{int, g}^m\hat{\boldsymbol{x}}_k-\sigma_k^2+\beta_{p,g}^m-\varphi_{int,g}^m(\xi_{\boldsymbol{h}_k}^2+\xi_{\boldsymbol{d}_k}^2)$, and $\boldsymbol{\varphi}_{int}^m=[\varphi_{int,1}^m,\cdots,\varphi_{int,G}^m]\geq 0$ are slack variables. In addition, $\tilde{\boldsymbol{\Xi}}_{int,g}^m=\tilde{\boldsymbol{\Psi}}_{int,g}^m(\tilde{\boldsymbol{\Psi}}_{int,g}^{m,(t)})^H+\tilde{\boldsymbol{\Psi}}_{int,g}^{m,(t)}(\tilde{\boldsymbol{\Psi}}_{int,g}^{m})^H-\tilde{\boldsymbol{\Psi}}_{int,g}^{m,(t)}(\tilde{\boldsymbol{\Psi}}_{int,g}^{m,(t)})^H$, where $\tilde{\boldsymbol{\Psi}}_{int,g}^m=[\boldsymbol{\Phi}\boldsymbol{H}\boldsymbol{W}_{-g}^m;\boldsymbol{I}_{M}]$ and $\tilde{\boldsymbol{\Psi}}_{int,g}^{m,(t)}=[\boldsymbol{\Phi}^{(t)}\boldsymbol{H}\boldsymbol{W}_{-g}^m;\boldsymbol{I}_{M}]$ and $\boldsymbol{W}_{-g}^m=[\boldsymbol{w}_{p,1}^m,\cdots,\boldsymbol{w}_{p,g-1}^m,\boldsymbol{w}_{p,g+1}^m,\cdots,\boldsymbol{w}_{p,G}^m,\boldsymbol{w}_{p,1}^u,\cdots,\boldsymbol{w}_{p,K}^u]$. LMI $\mathcal{F}_{ub}(\cdot)$ shares the same structure as $\mathcal{L}_{ub}(\cdot)$ but is applied to RIS phase optimization, where the variable is vector-valued.

With \eqref{eq34f} and \eqref{eq34g}convexified, the same methodology is applied to \eqref{eq34i}, \eqref{eq34j}, \eqref{eq34l} and \eqref{eq34m}. The resulting formulations are presented below, respectively
\begin{itemize}
	\item For \eqref{eq34i}: $\mathcal{F}_{lb}(\boldsymbol{\Xi}_{p,k}^u,\varphi_{p,k}^u,\delta_{p,k}^u)\succeq\boldsymbol{0}$, where $\delta_{p,k}^u = \hat{\boldsymbol{x}}_k^H \boldsymbol{\Xi}_{p,k}^u \hat{\boldsymbol{x}}_k - \alpha_{p,k}^u - \varphi_{p,k}^u (\xi_{\boldsymbol{h}_k}^2 + \xi_{\boldsymbol{d}_k}^2), \boldsymbol{\Xi}_{p,k}^u=\boldsymbol{\Psi}_{p,k}^u(\boldsymbol{\Psi}_{p,k}^{u,(t)})^H+\boldsymbol{\Psi}_{p,k}^{u,(t)}(\boldsymbol{\Psi}_{p,k}^{u})^H-\boldsymbol{\Psi}_{p,k}^{u,(t)}(\boldsymbol{\Psi}_{p,k}^{u,(t)})^H$, where $\boldsymbol{\Psi}_{p,k}^u=[\boldsymbol{\Phi}\boldsymbol{H}\boldsymbol{w}_{p,k}^u;\boldsymbol{I}_{M}]$ and $\boldsymbol{\Psi}_{p,k}^{u,(t)}=[\boldsymbol{\Phi}^{(t)}\boldsymbol{H}\boldsymbol{w}_{p,k}^u;\boldsymbol{I}_{M}]$.
	\item For \eqref{eq34j}: $\mathcal{F}_{ub}(\tilde{\boldsymbol{\Xi}}_{int,k}^u,\varphi_{int,k}^u,\delta_{int,k}^u)\succeq\boldsymbol{0}$, where $\delta_{int,k}^u = -\hat{\boldsymbol{x}}_k^H \tilde{\boldsymbol{\Xi}}_{int,k}^u \hat{\boldsymbol{x}}_k - \sigma_k^2 + \beta_{p,k}^u - \varphi_{int,k}^u (\xi_{\boldsymbol{h}_k}^2 + \xi_{\boldsymbol{d}_k}^2)$, $\tilde{\boldsymbol{\Xi}}_{int,k}^u=\tilde{\boldsymbol{\Psi}}_{int,k}^u(\tilde{\boldsymbol{\Psi}}_{int,k}^{u,(t)})^H+\tilde{\boldsymbol{\Psi}}_{int,k}^{u,(t)}(\tilde{\boldsymbol{\Psi}}_{int,k}^{u})^H-\tilde{\boldsymbol{\Psi}}_{int,k}^{u,(t)}(\tilde{\boldsymbol{\Psi}}_{int,k}^{u,(t)})^H$, where $\tilde{\boldsymbol{\Psi}}_{int,k}^u=[\boldsymbol{\Phi}\boldsymbol{H}\boldsymbol{W}_{-g,-k}^u;\boldsymbol{I}_{M}]$ and $\tilde{\boldsymbol{\Psi}}_{int,k}^{u,(t)}=[\boldsymbol{\Phi}^{(t)}\boldsymbol{H}\boldsymbol{W}_{-g,-k}^u;\boldsymbol{I}_{M}]$ and $\boldsymbol{W}_{-g,-k}^u=[\boldsymbol{w}_{p,1}^m,\cdots,\boldsymbol{w}_{p,g-1}^m,\boldsymbol{w}_{p,g+1}^m,\cdots,\boldsymbol{w}_{p,G}^m,\boldsymbol{w}_{p,1}^u,\cdots,\boldsymbol{w}_{p,k-1}^u,\\
	\boldsymbol{w}_{p,k+1}^u,\cdots,\boldsymbol{w}_{p,K}^u]$.
	\item For \eqref{eq34l}: $\mathcal{F}_{lb}(\boldsymbol{\Xi}_c,\varphi_{c,k},\delta_{c,k})\succeq\boldsymbol{0}$, where $\delta_{c,k} = \hat{\boldsymbol{x}}_k^H \boldsymbol{\Xi}_c \hat{\boldsymbol{x}}_k - \alpha_{c,k} - \varphi_{c,k} (\xi_{\boldsymbol{h}_k}^2 + \xi_{\boldsymbol{d}_k}^2)$, $\boldsymbol{\Xi}_{c}=\boldsymbol{\Psi}_c(\boldsymbol{\Psi}_c^{(t)})^H+\boldsymbol{\Psi}_{c}^{(t)}\boldsymbol{\Psi}_{c}^{H}-\boldsymbol{\Psi}_{c}^{(t)}(\boldsymbol{\Psi}_{c}^{(t)})^H$, where $\boldsymbol{\Psi}_{c}=[\boldsymbol{\Phi}\boldsymbol{H}\boldsymbol{w}_c;\boldsymbol{I}_{M}]$ and $\boldsymbol{\Psi}_{c}^{(t)}=[\boldsymbol{\Phi}^{(t)}\boldsymbol{H}\boldsymbol{w}_c;\boldsymbol{I}_{M}]$.
	\item For \eqref{eq34m}: $\mathcal{F}_{ub}(\tilde{\boldsymbol{\Xi}}_{int}^c,\varphi_{int,k}^c,\delta_{int,k}^c)\succeq\boldsymbol{0}$, where $\delta_{int,k}^c = -\hat{\boldsymbol{x}}_k^H \tilde{\boldsymbol{\Xi}}_{int}^c \hat{\boldsymbol{x}}_k - \sigma_k^2 + \beta_{c,k} - \varphi_{int,k}^c (\xi_{\boldsymbol{h}_k}^2 + \xi_{\boldsymbol{d}_k}^2)$, $\tilde{\boldsymbol{\Xi}}_{int}^c=\tilde{\boldsymbol{\Psi}}_{int}^c(\tilde{\boldsymbol{\Psi}}_{int}^{c,(t)})^H+\tilde{\boldsymbol{\Psi}}_{ini}^{c,(t)}(\tilde{\boldsymbol{\Psi}}_{ini}^{c})^H-\tilde{\boldsymbol{\Psi}}_{ini}^{c,(t)}(\tilde{\boldsymbol{\Psi}}_{ini}^{c,(t)})^H$, where $\tilde{\boldsymbol{\Psi}}_{int}^c=[\boldsymbol{\Phi}\boldsymbol{H}\boldsymbol{W}_{-c};\boldsymbol{I}_{M}]$ and $\tilde{\boldsymbol{\Psi}}_{int}^{c,(t)}=[\boldsymbol{\Phi}^{(t)}\boldsymbol{H}\boldsymbol{W}_{-c};\boldsymbol{I}_{M}]$ and $\boldsymbol{W}_{-c}^u=[\boldsymbol{w}_{p,1}^m,\cdots,\boldsymbol{w}_{p,G}^m,\boldsymbol{w}_{p,1}^u,\cdots,\boldsymbol{w}_{p,K}^u]$.
\end{itemize}

Following the preceding derivations, we adopt a penalty method similar to that used in $\widetilde{\mathcal{P}}_4$, which yields the following convex subproblem for optimizing $\boldsymbol{\phi}$: 
\begin{equation}
	\begin{aligned}
		\mathcal{P}_8:
		\max_{\mathcal{S}_2} &\sum\limits_{k\in\mathcal{K}} \left(C_{u,k}^u+\log_2(1+\gamma_{p,k}^u)\right)\notag\\
		&+\sum\limits_{g\in\mathcal{G}}\left(C_{c,g}^m+\log_2(1+\gamma_{p,g}^m)\right)\notag\\
		&+\mu\sum_{n\in\mathcal{N}}(2\Re\{(\phi_n^{(t)})^*\phi_n\}-\vert\phi_n^t\vert^2-1)\\
		\text { st. } &\eqref{eq34c}-\eqref{eq34e},\eqref{eq40},\eqref{eq43},\eqref{eq44},\\
		&\mathcal{F}_{lb}(\boldsymbol{\Xi}_{p,k}^u,\varphi_{p,k}^u,\delta_{p,k}^u)\succeq\boldsymbol{0},\\
		&\mathcal{F}_{ub}(\tilde{\boldsymbol{\Xi}}_{int,k}^u,\varphi_{int,k}^u,\delta_{int,k}^u)\succeq\boldsymbol{0},\\
		&\mathcal{F}_{lb}(\boldsymbol{\Xi}_c,\varphi_{c,k},\delta_{c,k})\succeq\boldsymbol{0},\\
		&\mathcal{F}_{ub}(\tilde{\boldsymbol{\Xi}}_{int}^c,\varphi_{int,k}^c,\delta_{int,k}^c)\succeq\boldsymbol{0},\\
		&f(\gamma_{p,k}^{u},\beta_{p,k}^{u},\alpha_{p,k}^{u})\geq0,f(\gamma_{c},\beta_{p,k},\alpha_{p,k})\geq 0,\\
		&\left|\phi_n\right|^2\leq 1, \forall n,
	\end{aligned}
\end{equation}
where $\mathcal{S}_2\triangleq\{\boldsymbol{\phi},\boldsymbol{\gamma}_{p}^m,\boldsymbol{\gamma}_{p}^u,\gamma_c,\boldsymbol{\alpha}_p^m,\boldsymbol{\alpha}_p^u,\boldsymbol{\alpha}_c,\boldsymbol{\beta}_p^m,\boldsymbol{\beta}_p^u,\boldsymbol{\beta}_c,\boldsymbol{\varphi}_{p}^m,\boldsymbol{\varphi}_{int}^m,\\
\boldsymbol{\varphi}_{p}^u,\boldsymbol{\varphi}_{int}^u,\boldsymbol{\varphi}_{c},\boldsymbol{\varphi}_{int}^c\}$. $\mathcal{P}_8$ can be solved by applying CVX tool with mosek toolbox. 

Finally, the overall BCA algorithm to solve $\mathcal{P}_6$ is summarized in \textbf{Algorithm \ref{Alg2}}.
\begin{algorithm}[htb]
	\caption{Robust beamforming design for CCF-RSMA.}
	\label{Alg2}
	\begin{algorithmic}[1]
		\STATE Initialize feasible sets $\{\boldsymbol{W}^{(0)},\boldsymbol{\phi}^{(0)},\boldsymbol{\gamma}_{p}^{m,(0)},
		\boldsymbol{\gamma}_{p}^{u,(0)},\boldsymbol{\gamma}_c^{(0)},$
		$\boldsymbol{\alpha}_p^{m,(0)},\boldsymbol{\alpha}_p^{u,(0)},\boldsymbol{\alpha}_c^{(0)},\boldsymbol{\beta}_p^{m,(0)},\boldsymbol{\beta}_p^{u,(0)},\boldsymbol{\beta}_c^{(0)}\}
		$, $\varepsilon^{(0)}=0$, penalty factor $\mu$, iteration number $t=0$. 
		\REPEAT
		\STATE For given $\{\boldsymbol{\phi}^{(t)},\varepsilon^{(t)}\}$, update the optimal $\{\boldsymbol{W}^{(t+1)},\boldsymbol{C}^{(t+1)}\}$ by solving $\mathcal{P}_7$.
		\STATE For given $\{\boldsymbol{W}^{(t+1)},\boldsymbol{C}^{(t+1)},\varepsilon^{(t)}\}$, update the optimal $\{\boldsymbol{\phi}^{(t+1)}\}$ by solving $\mathcal{P}_8$.
		\STATE Update $,\varepsilon^{t+1}$ based on \eqref{eq23}.
		\STATE $t=t+1$.
		\UNTIL{ the objective function of $\mathcal{P}_6$ is convergent. }
		\STATE Output $\{\boldsymbol{W}^{(t+1)},\boldsymbol{C}^{(t+1)},\boldsymbol{\phi}^{(t+1)}\}$.  
	\end{algorithmic}
\end{algorithm}

The EE maximization problem for PCF-RSMA under imperfect CSI shares the same structural form as $\mathcal{P}_5$ in Section III-B, but is defined over the channel with bounded uncertainty, as given in \eqref{eq33}. Since the subsequent robust reformulation and solution procedure follow the same methodology as developed for the CCF-RSMA scheme, we again omit the detailed derivation for brevity.
\section{Complexity Analysis}
Both Algorithm 1 and Algorithm 2 can be efficiently solved by interior-point method. For Algorithm 1, the number of variables and constraints for the $\{\boldsymbol{W},\boldsymbol{C}\}$-update are $a_1=(2M+3)(1+G+K)+K-2$ and $a_2=8K+2G+2$, respectively, while for the $\boldsymbol{\phi}$-update, they are $a_3=2N+2G+3K+1$ and $a_4=N+7K+G+1$, respectively. Hence, the per-iteration computational complexity is approximately $\mathcal{O}((a_1+a_2)^{3.5}\log_2(1/\varepsilon)+(a_3+a_4)^{3.5}\log_2(1/\varepsilon))$, where $\varepsilon$ is the convergence accuracy of interior point method. For Algorithm 2, the active beamforming design involves $a_5=2(M+N)^2(1+G+K)+10K+6G+1$ variables and $a_6=6K(M+N+1)^2+4(1+G+K)$, constraints, and the passive beamforming design involves $a_7=2N+2(G+K+1)+2$ variables and $a_8=N+6K(M+N+1)^2+2(G+K+1)$ constraints, leading to a per-iteration complexity of $\mathcal{O}((a_5+a_6)^{3.5}\log_2(1/\varepsilon)+(a_7+a_8)^{3.5}\log_2(1/\varepsilon))$.

Similarly, for the PCF-RSMA scheme under perfect CSI, the corresponding parameters are  $\acute{a}_1=(2M+1)(1+G+K)+3G+3K$, $\acute{a}_2=8K+4G+2$, $\acute{a}_3=2N+2G+3K+1$ and $\acute{a}_4=N+7K+G+1$, giving a complexity of $\mathcal{O}((\acute{a}_1+\acute{a}_2)^{3.5}\log_2(1/\varepsilon)+(\acute{a}_3+\acute{a}_4)^{3.5}\log_2(1/\varepsilon))$ Under imperfect CSI, the parameters become $\acute{a}_5=2(M+N)^2(1+G+K)+10K+7G+1$, $\acute{a}_6=6K(M+N+1)^2+5G+4K+4$, $\acute{a}_7=2N+2(G+K+1)+2$ and $\acute{a}_8=N+6K(M+N+1)^2+2(G+K+1)$, resulting in a complexity of $\mathcal{O}((\acute{a}_5+\acute{a}_6)^{3.5}\log_2(1/\varepsilon)+(\acute{a}_7+\acute{a}_8)^{3.5}\log_2(1/\varepsilon))$. Here, $\varepsilon$ denotes the convergence accuracy of the interior-point method.
\section{Simulation results}
\begin{table}[h]
	\caption{SIMULATION PARAMETERS}
	\centering
	\setlength{\tabcolsep}{2.5mm}{
		\begin{tabular}{ccc}\\
			\toprule [1pt] 
			Parameters & Values  \\
			\midrule
			Carrier frequency                        & $28$ GHz \\
			Bandwidth                                & $100$ MHz\\
			Noise power                              & $\sigma^2=-94$ dBm  \\ 
			Multicast rate requirement               & $R_m^{th}=1$ bps/Hz\\
			Unicast rate requirement                 & $R_u^{th}=0.1$ bps/Hz\\
			Power consumption of the baseband unit   & $P_{BB}=200$ mW\cite{intro29}\\
			Power consumption of the RF chain        & $P_{RF}=160$ mW\cite{intro29}\\
			Power consumption of the phase shifter   & $P_{PS}=20$ mW\cite{intro29}\\
			Power consumption of the power amplifier & $P_{PA}=40$ mW\cite{intro29}\\
			Power consumption of the RIS             & $P_{RIS}=20$ dBm\cite{intro28}\\
			Power consumption of each user           & $P_{U}=10$ dBm\cite{ref2}\\
			Power amplifier efficiency               & $\eta=0.38$\cite{intro28}\\
			Penalty factor                           & $\mu=100$\\
			Precision threshold                      & $0.001$\\
			\bottomrule
	\end{tabular}}
	\label{table2}  
\end{table}

This section evaluates the performance of the proposed algorithm via numerical simulations. In the simulation, we set the BS at $(0, 0, 10)$ m and the RIS at $(10, 10, 15)$ m. Users are uniformly distributed within a circle of radius $10$ m centered at $(80, 0, 1.5)$ m. The direct channel consists of three NLoS paths, while the cascaded channel comprises one LoS path and five NLoS paths. For each path, the elevation angle $\theta$ is uniformly distributed in $[0, \pi/2]$, and the azimuth angle $\phi$ is uniformly distributed in $[0, \pi]$. The complex gains of the channel paths, denoted by $\alpha_{k,l}$ for direct links and $\beta_{l}$ and $\beta_{k,l}$ for cascaded links, follow $\mathcal{CN}(0, 10^{-0.1 \cdot PL(d)})$, where $PL(d)$ denotes the distance-dependent path loss. Specifically, the path loss is modeled as $PL(d) = a + 10b\log_{10}(d) + \xi$, where $\xi \sim \mathcal{N}(0, \sigma_{\xi}^2)$ represents shadow fading. For the LoS path, we set $a = 61.4$, $b = 2$, and $\sigma_{\xi} = 5.8$ dB; for the NLoS paths, the parameters are $a = 72$, $b = 2.92$, and $\sigma_{\xi} = 8.7$ dB\cite{intro27}. The antenna gains are: $\lambda_B = 9.82$ dBi for the BS, $\lambda_U = 0$ dBi for users, and the RIS gain is given by $\frac{\lambda_R}{\sqrt{\lambda_B\lambda_U}} = 10$ dB\cite{intro28}. Other key simulation parameters are listed in Table \ref{table2}. All simulation results are obtained by averaging over $100$ independent channel realizations.

To demonstrate the effectiveness of the proposed CCF-RSMA and PCF-RSMA schemes, we compare them with four benchmark schemes, all optimized under the identical SCA-BCA algorithmic framework to ensure fairness:
\begin{enumerate}
	\item NoRIS: A baseline without RIS deployment, utilizing only the direct BS-user links.
	\item SDMA: All unicast and multicast streams are transmitted simultaneously via beamforming, and all interference is treated as noise.
	\item LDM-NOMA: Multicast and unicast streams are superimposed in the power domain, with multicast allocated higher power and decoded first via SIC\cite{intro6}.
	\item 2-layer Unsplitted-Multicast RSMA (UM-RSMA): This benchmark scheme adopts the CCF-RSMA scheme but keeps multicast messages complete\cite{intro11}. Unicast messags $W_k^u$ are still split into common parts $W_{c,k}^u$ and private parts $W_{p,k}^u$. All unicast common parts $\{W_{c,1}^u,\cdots,W_{c,K}^u\}$ are jointly encoded into a common stream $s_c$, while each multicast message $W_g^m$ and unicast private part are independently encoded into their respective stream $s_g^m$ and $s_{p,k}^u$. The two-layer structure is preserved to highlight the performance gain achieved by multicast splitting.
\end{enumerate}

For the bounded CSI error model, the relative amount of CSI uncertainties is measured by $\delta_{\boldsymbol{h}}\in[0,1)$ and $\delta_{\boldsymbol{d}}\in[0,1)$, which define the error variances as $\varepsilon^2_{\boldsymbol{h}_k}=\delta_{\boldsymbol{h}}^2\Vert\hat{\boldsymbol{h}_k}\Vert_2^2$ and $\varepsilon^2_{\boldsymbol{d}_k}=\delta_{\boldsymbol{d}}^2\Vert\hat{\boldsymbol{d}_k}\Vert_2^2$, respectively. Subsequently, the radii of the bounded uncertainty regions are given by $\xi_{\boldsymbol{h}_k} =\sqrt{\frac{\varepsilon_{\boldsymbol{h}_k}^2}{2} F_{2 N}^{-1}\left(1-\rho\right)}$ and $\xi_{\boldsymbol{d}_k} =\sqrt{\frac{\varepsilon_{\boldsymbol{d}_k}^2}{2} F_{2 M}^{-1}\left(1-\rho\right)}$, where $F_{2 N}^{-1}(\cdot)$ and $F_{2 M}^{-1}(\cdot)$ denotes the inverse cumulative distribution function of the Chi-square distribution with $2N$ and $2M$ degree of freedom, respectively. In the simulations, we set $\rho=0.05$ to define the uncertainty region\cite{alg1}.

\begin{figure}[h]   
	\begin{minipage}{0.49\linewidth}
		\vspace{3pt}  
		\centerline{\includegraphics[width=\textwidth]{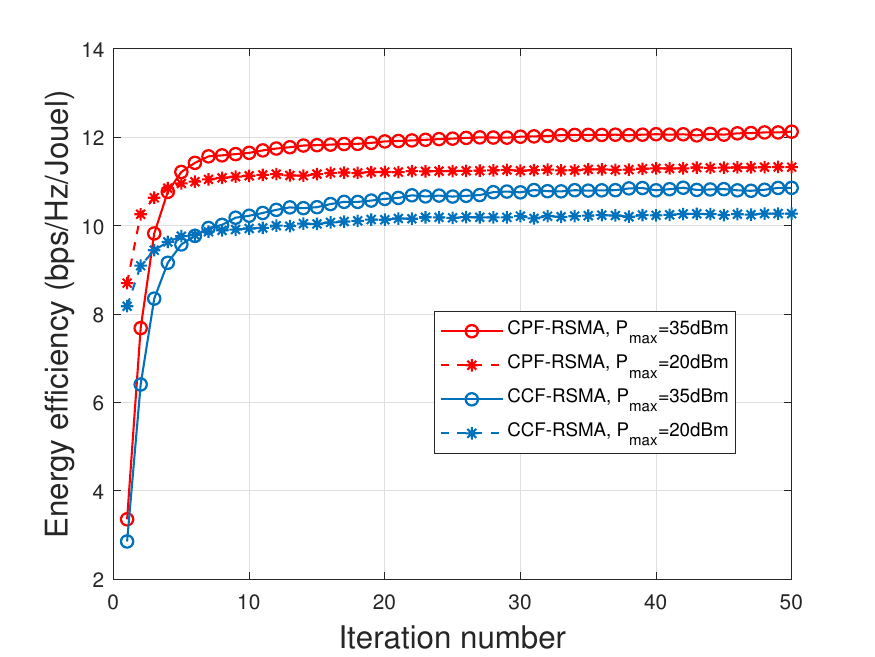}}
		\centerline{(a) Perfect CSI.}
	\end{minipage}
	\begin{minipage}{0.49\linewidth}
		\vspace{3pt}  
		\centerline{\includegraphics[width=\textwidth]{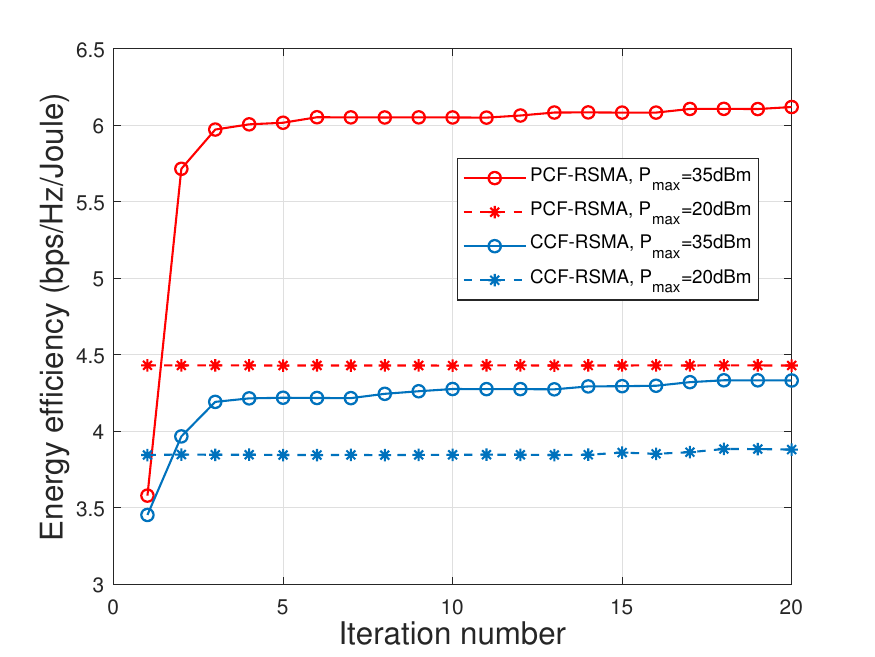}}
		\centerline{(b) Imperfect CSI.}
	\end{minipage}
	\caption{ \small{Convergence behavior of the proposed algorithms. (a) $M=36, N=64, K=8, G=4.$ (b) $M=9, N=12, K=4, G=2$, $\delta_{\boldsymbol{h}}=0.02$, $\delta_{\boldsymbol{d}}=0.02$.}}
	\label{fig4}
\end{figure}
Fig. \ref{fig4} illustrates the convergence behavior of the proposed algorithms. From Fig. \ref{fig4}(a), it can be seen that under perfect CSI, both PCF-RSMA and CCF-RSMA achieve fast and smooth convergence within $10–20$ iterations. Under imperfect CSI, as shown in Fig. \ref{fig4}(b), convergence is also achieved within $20$ iterations for all schemes, confirming robustness. At $P_{max}=20$dBm, the EE remains nearly constant from the first iteration with minor fluctuations, indicating that the limited power budget forces immediate near-optimal operation. At $P_{max}=35$dBm, a clear ascending trend is observed as the additional power enables progressive improvement of the solution. Overall, the proposed algorithms reliably converges under both CSI conditions, demonstrating its practical viability.

\begin{figure}[h]   
	\begin{minipage}{0.49\linewidth}
		\vspace{3pt}  
		\includegraphics[width=\linewidth]{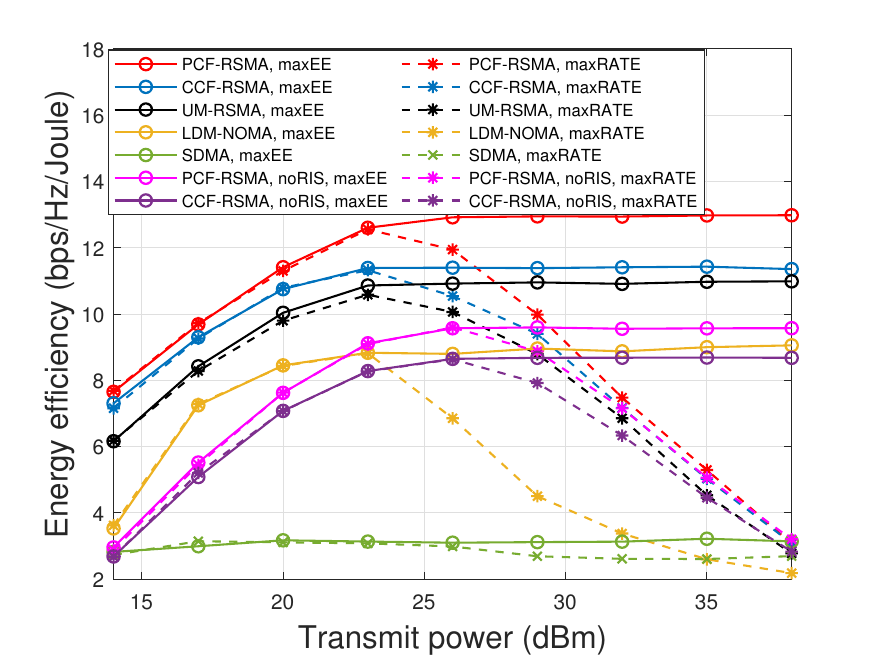}
		\centerline{(a)\scriptsize{M=36, N=64, K=8, G=4}.}
	\end{minipage}
	\begin{minipage}{0.49\linewidth}
		\vspace{3pt}  
		\centering
		\includegraphics[width=\linewidth]{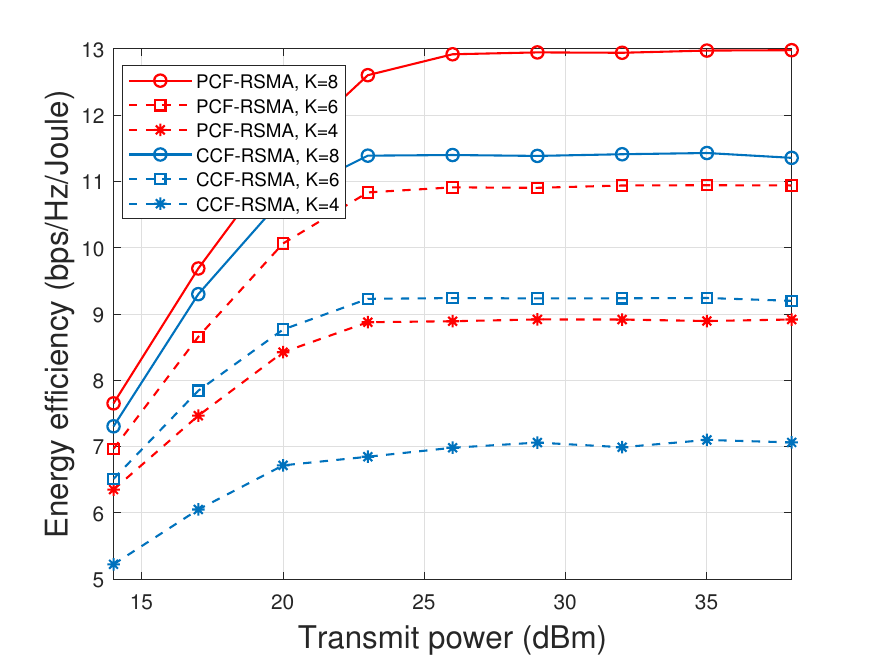}
		\centerline{(b)\scriptsize{M=36, N=64, K=\{4,6,8\}, G=\{2,3,4\}}.}
	\end{minipage}
	\begin{minipage}{0.49\linewidth}
		\vspace{3pt}  
		\includegraphics[width=\linewidth]{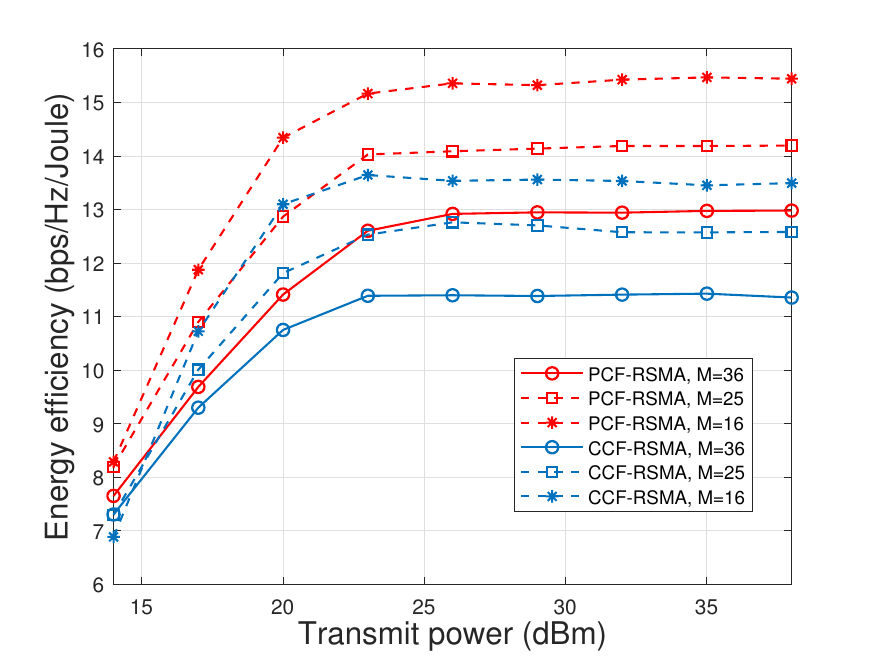}
		\centerline{(c)\scriptsize{N=64, K=8, G=4, M=\{25,36,64\}}.}
	\end{minipage}
	\begin{minipage}{0.49\linewidth}
		\vspace{3pt}  
	    \includegraphics[width=\linewidth]{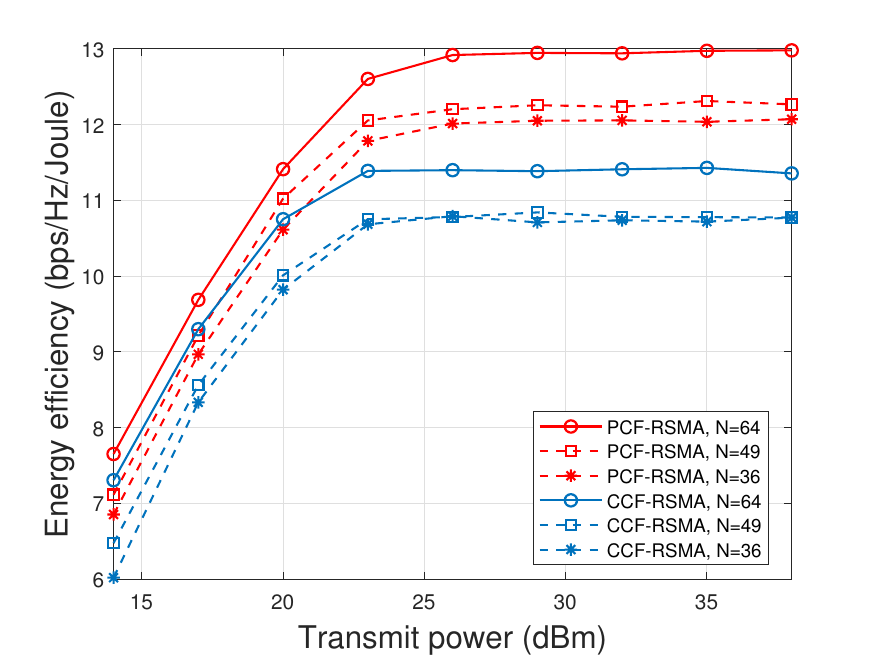}
	    \centerline{(d)\scriptsize{M=36, K=8, G=4, N=\{36,49,64\}}.}
	\end{minipage}
	\caption{\small{(a) Balance of rate and power. (b) EE versus user number. (c) EE versus Transmit antenna number. (d) EE versus RIS element number.}}
	\label{fig5}
\end{figure}

Fig. \ref{fig5} shows the EE performance under perfect CSI. Specifically, Fig. \ref{fig5}(a) demonstrates the EE versus transmit power under different transmission schemes. Two optimization objectives are considered. The first, denoted as ``maxEE", maximizes EE. The second, denoted as ``maxRATE", maximizes SE under the same constraints. For all schemes, EE increases with transmit power. At low power levels from $14$ to $23$ dBm, the EE achieved by ``maxEE" and ``maxRATE" are nearly identical for each scheme. This is because the limited power budget leaves no space for tradeoffs between the two objectives. As power increases, the two curves begin to diverge. The ``maxEE" curves continue to rise, while the ``maxRATE" curves saturate or even decline. This occurs because rate maximization allocates power aggressively to boost sum rate, but the additional rate gain from increasing transmit power diminishes at high power. Eventually, this additional gain fails to compensate for the increased power consumption, leading to a drop in EE.

﻿
Among the evaluated schemes, RSMA-based designs consistently outperform LDM-NOMA and SDMA. More importantly, both PCF-RSMA and CCF-RSMA, which split multicast information, achieve significantly higher EE than UM-RSMA. The gain is approximately $18.2\%$ and $3.4\%$, respectively, confirming the benefit of splitting multicast information. PCF-RSMA achieves the highest EE overall, followed closely by CCF-RSMA. The advantage of PCF-RSMA stems from its group-specific common streams, which enable finer-grained adaptation to intra-group channel conditions and better exploitation of multiuser diversity. LDM-NOMA outperforms SDMA but remains inferior to all RSMA variants. Finally, the presence of RIS significantly boosts EE for both PCF-RSMA and CCF-RSMA, with an improvement of approximately $35.6\%$ and $30.8\%$ compared to the noRIS cases.

Fig. \ref{fig5}(b) depicts the EE versus transmit power under different user/group configurations. First, the EE of both schemes increases consistently with $K$ and $G$, demonstrating their effectiveness across different user scales. Second, PCF-RSMA achieves significantly higher EE than CCF-RSMA in all cases, confirming its EE advantage and its ability to scale favorably with user/group density under the considered setup.

Fig. \ref{fig5}(c) depicts the EE versus the number of transmit antennas. The EE of both schemes decreases as the number of transmit antennas $M$ increases. This is because each antenna consumes a fixed circuit power in the system model. While more antennas provide beamforming gains that increase the sum rate, the rate improvement is outweighed by the growing circuit power consumption, leading to a decline in EE.

Fig. \ref{fig5}(d) illustrates the EE versus the number of RIS elements $N$. For both schemes, EE increases with $N$ due to higher passive beamforming gain. PCF-RSMA consistently outperforms CCF-RSMA across all $N$, as its group-specific common messages enable better adaptation to intra-group channels and more effective use of the passive beamforming gain from larger RIS arrays. In contrast, CCF-RSMA relies on a global common stream, which is less effective when each group has only two users and inter-group interference is weak. Consequently, its EE improvement with $N$ is limited, as its global coordination mechanism cannot fully utilize the additional spatial degrees of freedom from more RIS elements.

\begin{figure}[h]
	\begin{minipage}[t]{0.49\linewidth}	
		\centering
		\includegraphics[width=1\textwidth]{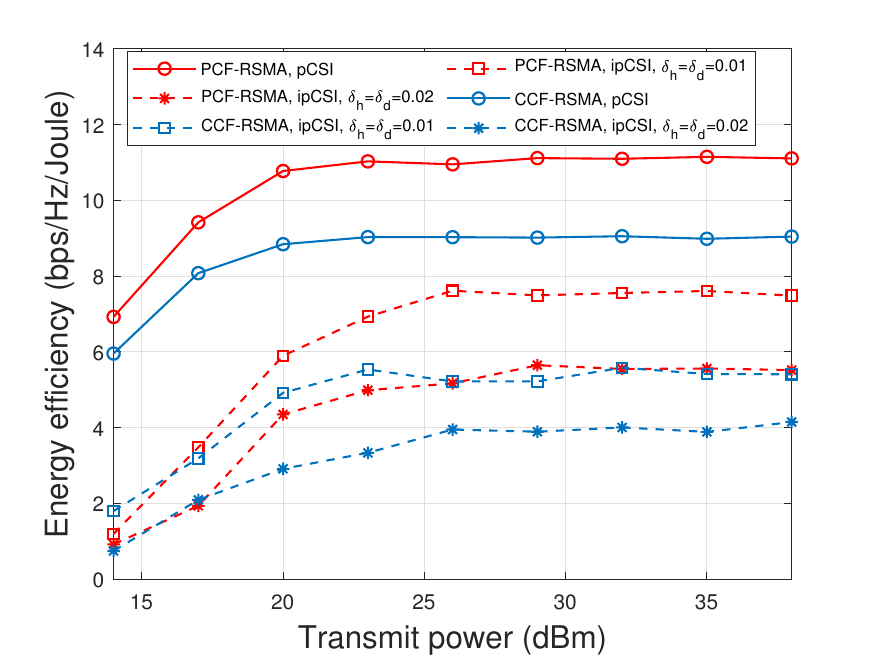}
		\caption{\small{EE versus transmit power under perfect and imperfect CSI, $M=9, N=12, K=4, G=2.$}}  
		\label{fig6}
	\end{minipage}
	\hfill
	\begin{minipage}[t]{0.49\linewidth}
		\centering
		\includegraphics[width=1\textwidth]{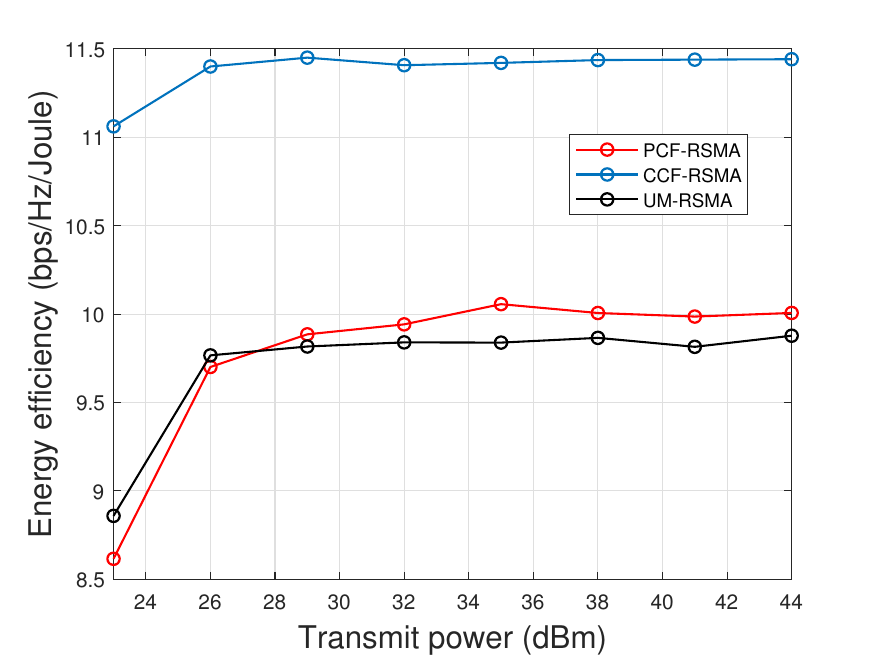}
		\caption{\small{EE versus transmit power, $M=9, N=12, K=8, G=2, R_m^{th}=0.1bps/Hz, R_u^{th}=1bps/Hz.$}}
		\label{fig7}
	\end{minipage}
\end{figure}

Fig. \ref{fig6} compares the EE under perfect CSI and imperfect CSI with different uncertainty levels. For both PCF-RSMA and CCF-RSMA, EE under perfect CSI is significantly higher than that under imperfect CSI, confirming the performance degradation caused by channel estimation errors. When the uncertainty level is reduced form $\delta=0.02$ to $\delta=0.01$, EE increases substantially indicating that the proposed robust design is sensitive to the accuracy of channel estimation. Under imperfect CSI with the same uncertainty level, PCF-RSMA achieves higher EE than CCF-RSMA across all transmit power levels, consistent with the trend observed in perfect CSI.

Fig. 7 shows the EE under a different configuration with $K=8$ users divided into $G=2$ multicast groups, i.e., 4 users per group, where the multicast QoS is $0.1$ bps/Hz and the unicast QoS is $1$ bps/Hz. In this setting, CCF-RSMA achieves higher EE than PCF-RSMA across all transmit power levels. This result contrasts with previous observations where PCF-RSMA outperformed CCF-RSMA under configurations with fewer users per group and higher multicast QoS requirements. The reason lies in the interplay between group size and QoS demands. When each group contains more users, the multicast common stream already serves a larger number of users. By merging the unicast common part into this global common stream, CCF-RSMA can convey additional information at low marginal cost. Meanwhile, the high unicast QoS requires substantial power for the unicast private streams. The power saved by CCF-RSMA through global stream sharing can be reallocated to support these demanding unicast streams. In contrast, PCF-RSMA places the unicast common part in the group-level stream, which cannot leverage the same sharing efficiency when groups are large, resulting in lower EE. These observations indicate that the optimal fusion scheme depends on both group size and the distribution of QoS requirements.

Importantly, both CCF-RSMA and PCF-RSMA consistently outperform UM-RSMA. This confirms that splitting multicast information provides tangible benefits regardless of which fusion scheme proves optimal for a given parameter configuration. These observations indicate that the optimal fusion scheme depends on both group size and the distribution of QoS requirements, while the advantage of multicast splitting itself remains consistent across different setups.

\section{CONCLUSION}
This paper investigates the EE optimization problem for joint unicast and multi-group multicast transmission in RIS-assisted mmWave networks. To address the complex interference in scenarios with coexisting mixed services, we propose two structured RSMA  schemes, i.e., CCF-RSMA and PCF-RSMA. For each scheme, we formulate a unified EE maximization problem that accounts for both perfect and imperfect CSI, and develop efficient joint optimization algorithms for both cases. Simulation results confirm the effectiveness of multicast information splitting, with both schemes consistently outperforming comparative schemes. Key insights reveal that the proposed framework is superior to the scheme without multicast splitting and the optimal RSMA scheme depends on system parameters. Specifically, CCF-RSMA is preferable when groups are larger with low multicast QoS, while PCF-RSMA exceeds when groups are smaller with high multicast QoS.

\section{ACKNOWLEDGEMENTS}
This work was supported by the National Key Research and Development Program of China under Grant 2021YFA071660.

\end{document}